# Distributed matching scheme and a flexible deterministic matching algorithm for arbitrary systems

Yu-Chiu Chao[*]

SLAC National Accelerator Laboratory, Menlo Park, California 94025 USA

We discuss the distributed matching scheme in accelerators where control of transverse beam phase space, oscillation, and transport is accomplished by flexible distribution of focusing elements beyond dedicated matching sections. Besides freeing accelerator design from fixed matching sections, such a scheme has many operational advantages, and enables fluid optics manipulation not possible in conventional schemes. Combined with an interpolation scheme this can bring about a new paradigm for efficient, flexible, and robust optics control. A rigorous and deterministic algorithm is developed for its realization. The beam phase space dynamics is naturally integrated into the algorithm, as opposed to being treated as generic numerical parameters as in traditional schemes, and thus better equipped to deal with physics motivated scenarios. The algorithm is a matching tool in its own right with unique characteristics in robustness and determinism. The beam phase space dynamics is naturally integrated into the algorithm, instead of being treated as generic numerical parameters as in traditional schemes. It is applicable to a wider range of problems, such as trading-off between competing options for desired machine states.

## I. INTRODUCTION

Standard paradigm for accelerator design and operation calls for dedicated matching sections tasked with locally controlling beam phase space, containing orbit jitter, or restoring optical transport properties at designated locations. Using a small fraction of focusing elements and usually spaced far apart, such sections are nonetheless expected to correct errors built up from far upstream, and ensure matched beam or transport over a long span downstream. In between are the remaining majority of focusing elements (quadrupoles and skew quadrupoles), deemed untouchable and passively holding up beam transport. Excluding esoteric cases, this paradigm came about mainly due to well-known machine tuning problems caused by indiscriminate use of focusing elements unaided by monitoring devices and competent algorithm. In this paradigm, the benefits otherwise available from these passive focusing elements remain untapped, much like the case of orbit correction using only a limited set of "allowed" correction magnets before real time beam position monitors, competent steering algorithms, and adequate computing power became a staple of accelerator operation.

In Section II we will discuss the benefits afforded by an alternative distributed matching scheme, in which the restriction of localized matching is lifted, using well-tested monitoring methods as inputs. In Section III a rigorous and deterministic algorithm is presented, which is a natural candidate for realizing distributed matching, as well as a robust and versatile matching algorithm in its own right. Nuances and extended application scenarios of the algorithm will also be discussed. In Section IV we will discuss how to integrate the schemes developed above into a deterministic, rigorous and efficient operating paradigm. In Section V we present applications of the algorithm in realistic accelerator situations.

The remainder of this report uses matching of uncoupled optics as numerical examples. But as is clear from the formulation to be introduced, the method is readily applicable to matching XY-coupled transport including skew quadrupoles.

## II. DISTRIBUTED MATCHING

### A. Local vs Distributed Matching

Distributed matching is motivated by the example illustrated in Figure 1 showing two scenarios of the beam envelope under a local (blue) and a distributed (red) matching scheme. The example is created with pronounced effects for emphasis but based on real optics. In both scenarios the beam envelope starts mismatched with sizeable beta-beat from the left. In the local scheme since most quadrupoles are not used for tuning, the correction has to wait until the dedicated matching section, while mismatch continues to build up. The local matching section usually consists of limited number of quadrupoles, not necessarily optimally configured for the accidental incoming mismatch, and as a result, can often result in drastic matching requiring excessive quadrupole strengths, significant local beam blow-up, or simply failure to get a complete match. The drastic solution itself increases the likelihood of matching failure with increased demand on measurement and implementation precision. Any such failure, whether algorithmic, numerical, or instrumental, will leak beyond the matching section, to which one has no recourse but to live with its consequences all the way until the next matching section.

---

[*] ychao@slac.stanford.edu Work supported in part by National Science & Engineering Research Council (NSERC) of Canada

Alternatively, Figure 1 shows how a distributed matching scheme can alleviate all the above problems by cutting down mismatch everywhere, achieving solutions more gentle on beam envelope and magnet strength, and dynamically counteracting matching failure wherever it happens. Such a scheme can be realized provided methods are available to effectively measure/monitor beam mismatch or transport error, as are algorithms to rigorously and unequivocally define and solve for partial matching solutions.

With the possibility to harness all quadrupoles for matching, one also gains a new degree of freedom in controlling how graduated matching is to be realized. For example, the rate of mismatch reduction along the beam line can be traded off against the degree of perturbation to existing optics. The choice made for the distributed matching of Figure 1 reflects a bias toward the latter. Such freedom can be highly prized in difficult operation situations, but is not an option when a designated local matching section is all there is for matching. Furthermore, by identifying "points of diminished return" in the algorithm introduced in Appendix B, a distributed matching scheme can get the best "bang for the buck" on all quadrupoles in a systematic way. No such freedom is possible with a localized matching scheme, which has to achieve complete matching, even if it means driving quadrupoles into highly ineffective regimes beyond points of diminished return.

Large beam envelope or jitter over extended areas as a result of localized matching in Figure 1 not only creates problems for machine control and hardware, but can adversely impact beam quality as such envelope or jitter can sample orbit dependent effects, such as nonlinearities or chromatic effects, in irreversible ways. Drastic matching can also result in small beam size and large divergence with space charge or other undesirable effects. It is to the best interest of accelerator performance to keep the beam envelope/ and divergence to design and jitter to minimum as gently and predictably as possible at all locations, not just points immediately after matching sections.

As an accelerator tuning activity, transverse phase space control is characterized by geographically prevalent distribution of actuators and monitors, error sources, and error-induced damage, all of which argue for a distributed correction scheme, much more akin to orbit correction than tuning with localized impact, error source, and actuator/monitor, such as bunch length or dispersion control. Absence of a robust, efficient, and intuitively simple algorithm may be what prevented its realization. But one should not lose sight of the fact that the primary role of all quadrupoles is envelope/jitter containment, preferably in an active sense, and thus the debate should not be about whether, but how, to use all quadrupoles for matching.

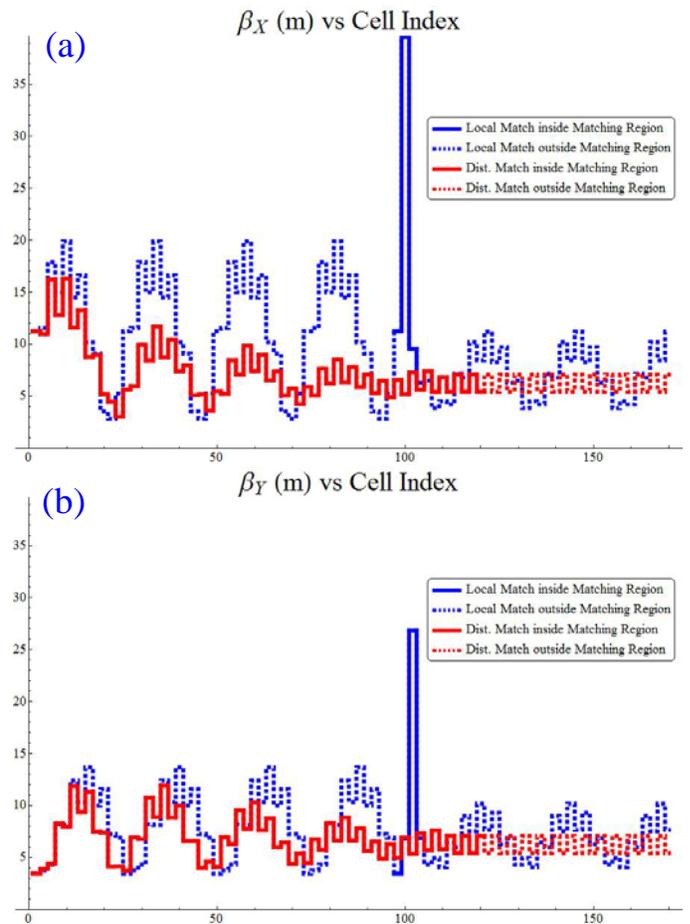

(a) $\beta_X$ (m) vs Cell Index

(b) $\beta_Y$ (m) vs Cell Index

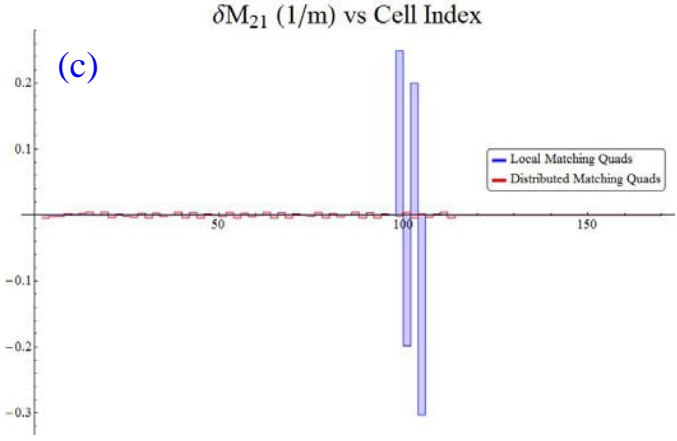

Figure 1. Physically realistic example with pronounced effects comparing matching under local (blue) and distributed (red) schemes in a 30° FODO Lattice. (a–b): βx and βy inside regions of matching quadrupoles (solid line) and outside (dashed line), (c): Changes in quadrupole inverse focal length needed to complete matching.

The appeal of distributed matching scheme notwithstanding, questions remain as to how to define the graduated profile for mismatch reduction of Figure 1 in a systematic and optimal way. This will be the focus of the following sections.

In a more ambitious scheme, matching can be made to resemble orbit correction even more closely, by demanding matching of beam Twiss parameters to design at multiple locations through an adequate ensemble of quadrupoles in the presence of local errors. It will be easier to elucidate and motivate this after further discussion of the deterministic algorithm introduced in the following sections, and will be deferred until Section V3.

### B. Measuring and Monitoring Mismatch and Transport Error

In this section we briefly define mismatch in the current context, and discuss well-tested methods to measure or monitor beam mismatch or transport error in the transverse plane, and terminology relevant to later presentation.

#### 1. Beam mismatch

Mismatch of measured beam profile is always cited against a design beam, as shown in Figure 2(a), where discrepancy between measured and design beams in the phase space can be quantified by a mismatch factor

$$\Phi = \left(\beta_D \cdot \gamma_M - 2\, \alpha_D \cdot \alpha_M + \gamma_D \cdot \beta_M\right)/2 \geq 1, \quad (1)$$

with equal sign corresponding to fully matched beam. Subscripts $M$ and $D$ denote measured and design beams, and $\alpha$, $\beta$, $\gamma$ are their Twiss parameters in either X or Y plane. Operationally $\Phi$ is computed by measuring Twiss parameters of the beam through multiple beam profile monitors such as screens or wire scanners, with the option of controlled optical variations to sample different cross sections of the phase space, often referred to as quadrupole scans.

#### 2. Transport error

Effects of machine transport deviation from design can be analogously characterized by a mismatch factor. The purpose of correcting transport error is to restore the beam to matched state at the exit of the region of interest assuming matched state at the entrance. In this respect it is different from matching the beam phase space, as here one only pays attention to the properties of the transport section. Figure 2(b) illustrates the effect of transport error on an originally matched beam, and outcome of restored transport. Transport error can be monitored or measured most efficiently with difference orbit methods, where beam trajectory response to an induced kick is measured and the following quantity computed at all position monitors:

$$C = \beta_D \cdot \delta X'^2 + 2\, \alpha_D \cdot \delta X \cdot \delta X' + \gamma_D \cdot \delta X^2 \quad (2)$$

where $\delta X$ and $\delta X'$ are deviations from un-kicked position and angle of the beam trajectory at a given location. $C$ is invariant so long as there is no transport error, and undergoes changes when errors are encountered. In Figure 2(b) the blue (red) dot represents the design (measured) beam trajectory in phase space. Clearly restoring the invariance of $C$ in equation (2) for a single trajectory is a necessary but not sufficient condition for fixing the transport error. To do the latter $C$ must be made invariant for beam trajectory of arbitrary phases, or equivalently, for the entire ellipse traced out by trajectories by all induced kicks. To capture this condition, we can introduce an extended version of the mismatch factor of equation (1)

$$\Phi = Tr\left(\Sigma_D^{-1} \cdot \Sigma_M\right)/(2N) \quad (3)$$

where $N$ is the dimensionality of the phase space, and $\Sigma_{D/M}$ are the covariance matrices of all kicked trajectories under design and measured transports respectively ($\delta$ is dropped in $\delta X$ and $\delta X'$ below for compactness),

$$\Sigma_{D/M} = \begin{pmatrix} \langle X_{D/M}^{1} \cdot X_{D/M}^{1} \rangle & \langle X_{D/M}^{1} \cdot X_{D/M}^{\prime 1} \rangle & \cdots & \langle X_{D/M}^{1} \cdot X_{D/M}^{N} \rangle & \langle X_{D/M}^{1} \cdot X_{D/M}^{\prime N} \rangle \\ \langle X_{D/M}^{\prime 1} \cdot X_{D/M}^{1} \rangle & \langle X_{D/M}^{\prime 1} \cdot X_{D/M}^{\prime 1} \rangle & \cdots & \langle X_{D/M}^{\prime 1} \cdot X_{D/M}^{N} \rangle & \langle X_{D/M}^{\prime 1} \cdot X_{D/M}^{\prime N} \rangle \\ \vdots & \vdots & \ddots & \vdots & \vdots \\ \langle X_{D/M}^{N} \cdot X_{D/M}^{1} \rangle & \langle X_{D/M}^{N} \cdot X_{D/M}^{\prime 1} \rangle & \cdots & \langle X_{D/M}^{N} \cdot X_{D/M}^{N} \rangle & \langle X_{D/M}^{N} \cdot X_{D/M}^{\prime N} \rangle \\ \langle X_{D/M}^{\prime N} \cdot X_{D/M}^{1} \rangle & \langle X_{D/M}^{\prime N} \cdot X_{D/M}^{\prime 1} \rangle & \cdots & \langle X_{D/M}^{\prime N} \cdot X_{D/M}^{N} \rangle & \langle X_{D/M}^{\prime N} \cdot X_{D/M}^{\prime N} \rangle \end{pmatrix}, \quad (4)$$

with angle brackets indicating ensemble averaging. When normalized in one dimension, $\Sigma_{D/M}$ are analogous to matrices formed by design and measured Twiss parameters, and with no loss of generality $\Sigma_{D/M}$ can be viewed as beam covariances if each difference orbit is replaced with the particle trajectory inside the beam relative to the centroid. Note as mentioned earlier, $\Phi$ of equation (3) is referred against a design transport. In other words, the trajectories of equation (4) are those at the exit of a transport section, whether design or measured, when the trajectory ensemble is matched to design at the entrance. This can be made more explicit through

$$\Sigma_{M}^{Out} = M(k_m) \cdot \Sigma_{D}^{In} \cdot M^{T}(k_m) \quad (5)$$

where *In/Out* denote covariances computed at the entrance/exit of the transport section of interest, and $M$ is its transfer matrix consisting of many optical elements of focusing strengths $k_m$. Equation (3) becomes

$$\Phi = Tr\left(\Sigma_{D}^{Out^{-1}} \cdot M(k_m) \cdot \Sigma_{D}^{In} \cdot M^{T}(k_m)\right) / (2N), \quad (6)$$

an expression involving only the design beam covariances and the transfer matrix, with no regard to measured beam profile. In Appendix A we will show the important result that $\Phi \geq 1$, and $\Phi = 1$ if and only if the design and measured beam covariances at the exit, $\Sigma_D$ and $\Sigma_M$, are identical, which is the goal of transport error correction[1].

Invocation and measurement of difference orbits of equations (2) and (4), as well as resulting determination of local transport errors implied by equation (6) through transfer matrices, are easily accessible operational tools at most accelerators [1,7].

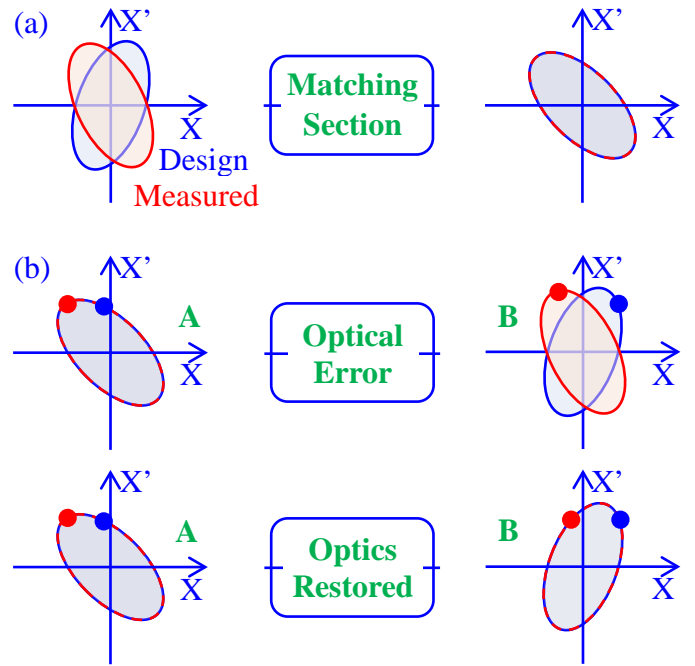

Figure 2. (a): Action of matching restores agreement between design and measured beam phase space characteristics. (b): Transport error causes original on-design beam to deviate from design (top), to be corrected by matching (bottom). This correction pertains only to the on-design input beam.

### C. Conventional Matching Algorithms

Conventionally the majority of matching problems are cast in the form of a numerical optimization problem with a penalty function constructed of a combination of individual Twiss parameter errors with respect to design values, and in many cases also of quadrupole strengths or other figures of merit. Powerful numerical optimization engines are brought to bear on such problems to arrive at a solution. This approach has competently met demands under many accelerator design and operation situations, and sees little incentive for change in such routines. In other situations, especially ones discussed in the current report, this approach can fall short, where the deterministic algorithm presented below can play a complementary role. The shortcomings of the conventional scheme stem from the fact that it is not informed by the underlying physics. It treats the problem as a pure numerical one, with all variables and constraints being generic numbers stripped of their physical meaning. As a result one can encounter the following difficulties:

- There is no physically unambiguous prescription to weigh between Twiss parameters α and β. This also applies to matching at more than one location.
- There is no unambiguous prescription to weigh between Twiss parameters, quad strength and other target parameters or control variables.
- There is no unequivocal criterion to determine if the best possible solution has been reached, in cases of multiple solutions, or in cases of no solution allowed by given configuration.
- There is no unambiguous prescription to extract partial solutions, or compare between different partial solutions.

---

[1] This does not mean the measured transfer matrix $M$ is identical to the design transfer matrix, which is a stronger requirement than ordinary transport error correction.

The above claim does not intend to be categorical. There may exist matching algorithms free of these difficulties, although it is less likely the case if matching is treated simply as a generic optimization problem not informed by underlying physics. The algorithm introduced in the following section, dictated by the physical significance of mismatch, proves more adept and consistent at addressing these issues.

### III. DETERMINISTIC MATCHING ALGORITHM

It is clear from Figure 1 that the essence of a distributed matching scheme is monotonically reduced mismatch $\Phi$, defined in equation (1), (3) or (6), over a continuous array of elements until $\Phi=1$. At every stage of reduction the scheme should provide an unequivocal recipe for taking $\Phi$ from one value to the next in a rigorously optimal way. Intermediately matched states imply degenerate solutions, namely, for an intermediate $\Phi>1$, the solution is not unique in terms of corresponding Twiss parameters, let alone quadrupole strengths $k_m$. This degeneracy is actually welcome, as it invites additional constraints to regain solution uniqueness through optimality requirements. This is hardly unfamiliar, as matching is never a single minded pursuit at the expense of all other factors such as quadrupole strengths or perturbation to existing optics. There is always a trade-off between reducing matching and other constraints, although it has not been put on a formal and rigorous footing. An extension to the method of Lagrange multipliers can provide just such a formulation. This is discussed in detail in Appendix B. We will outline its findings below.

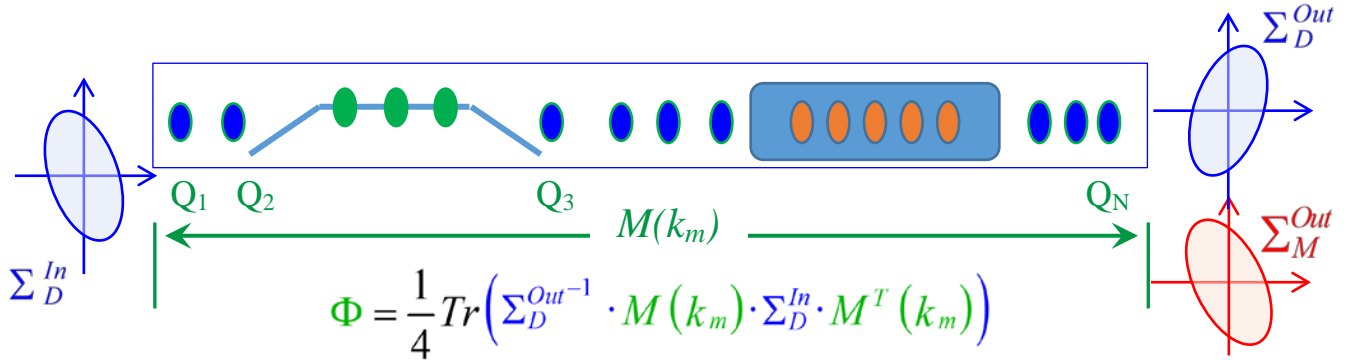

Figure 3. Concept of a generic matching problem, especially in terms of transport error correction. The matching section $M$ consists of quadrupoles $Q_1$, $Q_2$, …… $Q_N$, interleaved with other optical modules. In the presence of transport error, design beam covariance $\Sigma_D$ at the entrance to $M$ is translated into $\Sigma_M \neq \Sigma_D$ at the exit, which is equivalent to $\Phi>1$. The action of matching restores $\Phi=1$, and $\Sigma_M = \Sigma_D$.

#### A. Deterministic Matching Algorithm

*1. Competing objective H*

Figure 3 captures the essence of a matching problem with the goal of restoring the mismatch factor $\Phi$ defined in equation (6) to unity through changes in a total of $N_Q$ quadrupole with strengths $k_m$. The meaning of $\Phi=1$ is equivalent to the statement "If the beam covariance is matched to design at the entrance, then it is matched to design at the exit" (See Appendix A). As explained above, a competing objective is needed to constrain all intermediate solutions by the method of Lagrange multipliers. This competing objective can be taken from either the absolute quadrupole strengths,

$$H = \text{K}(\boldsymbol{k}) = \sum_{m=1}^{N_Q} k_m^2, \tag{7}$$

or deviation from a given set of baseline quadrupole strengths,

$$H = \Delta \text{K}(\boldsymbol{k}) = \sum_{m=1}^{N_Q} \delta k_m^2 = \sum_{m=1}^{N_Q} \left(k_m - k_m^D\right)^2, \tag{8}$$

where superscript $D$ represents *Design*.

### 2. The integration recipe

We recast the singularity-free recipe developed in Appendix B in the following, using as competing objectives $F=\Phi$ of (6) and $H=\text{K}$ of (7).

1. The goal is to obtain a continuous family of solutions in quadrupole strengths $k_m$ representing optimal trade-off between mismatch $\Phi(\boldsymbol{k})$ and quadruple strength $\text{K}(\boldsymbol{k})$ at every point, parametrized by a non-positive number $\mu = 1/\lambda$. Partial matching solutions can be taken from this 1D solution family, which always gives the best $\Phi$ for given K, and vice versa.
2. The process starts with initial values and slopes

$$\mu = 0, \qquad \left.\frac{d\mu}{dk}\right|_{\substack{\mu=0 \\ k_m=0}} = -\frac{2}{|\boldsymbol{T}|},$$

$$k_m = 0, \qquad \left.\frac{d\boldsymbol{k}}{dk}\right|_{\substack{\mu=0 \\ k_m=0}} = -\hat{\boldsymbol{T}}, \tag{9}$$

$$m = 1, 2, \ldots, N_Q, \qquad \boldsymbol{T} = \left.\nabla_{\boldsymbol{k}} \Phi(\boldsymbol{k})\right|_{k_m=0},$$

where we shorthanded $|d\boldsymbol{k}|$ as $dk$, bold faced letters $\boldsymbol{k}$ and $\boldsymbol{T}$ denote vectors, and caret denotes unit vector. The form of $d\boldsymbol{k}/dk$ above, including the minus sign, is intuitively clear from discussion in Appendix B.

3. The following formulas are used to solve for the trade-off curve by integration from the above starting point, with $\mu$ and the vector $\boldsymbol{k}$ as dependent variables and the length of $\boldsymbol{k}$, $|\boldsymbol{k}|$, as independent variable:

$$\frac{d\boldsymbol{k}}{dk} = -\hat{\boldsymbol{Q}}, \qquad \boldsymbol{Q} = Adj(\boldsymbol{N}) \cdot \boldsymbol{S},$$

$$\frac{d\mu}{dk} = -\frac{Det(\boldsymbol{N})}{|\boldsymbol{Q}|} \tag{10}$$

where bold faced letters $\boldsymbol{k}$ and $\boldsymbol{Q}$ denote vectors, and

$$N_{ij} = \frac{\partial^2 \left(\text{K}(\boldsymbol{k}) - \mu \cdot \Phi(\boldsymbol{k})\right)}{\partial k_i \partial k_j}, \qquad S_i = \frac{\partial \Phi(\boldsymbol{k})}{\partial k_i}, \tag{11}$$

with $Adj(\boldsymbol{N})$ the adjugate, or transpose of cofactor, of $\boldsymbol{N}$

$$Adj(\boldsymbol{N}) = Cof(\boldsymbol{N})^T = Det(\boldsymbol{N}) \cdot \boldsymbol{N}^{-1}. \tag{12}$$

4. As $\mu$ develops into a finite negative number, set $\lambda = 1/\mu$ and switch to integration formulas below, with $\lambda$ and the vector $\boldsymbol{k}$ as dependent variables and the length of $\boldsymbol{k}$, $|\boldsymbol{k}|$, as independent variable

$$\frac{d\boldsymbol{k}}{dk} = +\hat{\boldsymbol{P}}, \qquad \boldsymbol{P} = Adj(\boldsymbol{M}) \cdot \boldsymbol{R},$$

$$\frac{d\lambda}{dk} = +\frac{Det(\boldsymbol{M})}{|\boldsymbol{P}|} \tag{13}$$

where $\boldsymbol{P}$ denote vectors (note sign change), and

$$M_{ij} = \frac{\partial^2 \left(\Phi(\boldsymbol{k}) - \lambda \cdot \text{K}(\boldsymbol{k})\right)}{\partial k_i \partial k_j}, \qquad R_i = \frac{\partial \text{K}(\boldsymbol{k})}{\partial k_i}. \tag{14}$$

5. Do not stop as long as $\lambda \neq 0$, no matter how close $\Phi$ approaches unity.
6. Stop as soon as $\lambda=0$, even if $\Phi$ has not reached unity.

If $H=\Delta\text{K}$ of (8) is used instead of $H=\text{K}$ as the competing objective, make the following replacements in step 2, equation (9), and repeat the same recipe:

$$k_m = 0 \to k_m = k_m^D. \tag{15}$$

Nothing keeps one from using yet another competing objective $H$ different from (7) and (8), for which the above recipe is perfectly applicable as long as the underlying physics makes sense. One interesting possibility is to use as competing objective a different Φ, representing mismatch factor relative to a different design optics. In this case the trade-off will be between two competing optics. Its potential application will be explored in Section V.

Rationale and further detail of the above recipe are discussed in Appendix B. At the end of this procedure one obtains a one-dimensional path in the space spanned by quadrupole strengths $k_m$, each point of which represents an optimal trade-off between the mismatch (Φ) and quadrupole strength (K or ΔK) with a unique bias: Larger absolute value of $\lambda$ ($\mu$) favors more optimal Φ (K/ΔK). Note the procedure begins with ($\mu$=0; $\lambda$=−∞) and ends with ($\lambda$=0; $\mu$=−∞).

## B. Advantages over Conventional Matching Algorithm

The case for distributed matching has been argued in Section IIA. The algorithm introduced in the previous section is a perfect candidate for distributed matching, as it provides a rigorous recipe to realize optimal intermediate matching at every stage, regardless of the detail as to how the mismatch profile is tapered. Detail of this implementation will be discussed in Section IV.

Besides realizing distributed matching, the advantages of the algorithm as a pure matching tool are unique, marking a departure from conventional methods.

1. Determinism of the algorithm removes need for inspired guesses, random number search, or parameter tweaking typical of many convention matching methods. There is no free parameter in the algorithm requiring cases-by-case adjustment.
2. Determinism gives unequivocal criteria on when to stop and when not to stop the procedure. Conventional methods, lacking rigorous guidance, can fail on both criteria. The former instance of failure (stop on $\lambda$=0) often happens in 4-quadrupole matching with no solution [2], and the latter (do not stop on $\lambda$≠0) is illustrated in Appendix B2.
3. Removal of random and ad hoc parameter tweaking enables its usage in applications demanding smooth response to variation in fractional matching target, input mismatch, intervening transport optics, etc., and avoidance of case-by-case human intervention. Example applications include real time betatron matching feedback, and interpolated matching solution from pre-calculated database to be discussed in Section IV.
4. Robustness has been established through extensive tests, some of which are discussed in Appendix B. These include excessive initial mismatch (Φ>7000), and difficult matching configuration in low phase advance lattice (30° FODO) due to poorly-decoupled X & Y β-functions.
5. Good scalability is expected. As it depends on integration, or locally solving differential relations, as opposed to multiple objective optimization or root finding, the computation complexity has a weaker dependence on the size (e.g., number of quadrupoles) of the problem. As matching system becomes larger, use of conventional methods can be progressively more difficult. There is a definite advantage in using more than 4 quadrupoles for matching. Dedicated matching section and algorithmic/numerical complication have largely discouraged such practice. The current algorithm applied to distributed matching should remove these obstacles.
6. The algorithm gives a rigorous definition of optimal partial matching solution, as opposed to frequently ill-defined partial solution recipes in conventional methods.
7. The algorithm is readily adaptable to more complicated configurations, such as full XY-coupled matching, other physically sound matching constraints, or matching quadrupoles interspersed with nontrivial, special purpose optics.
8. There is no need for ad hoc weighted merit function between incongruent Twiss parameters and quadrupole strengths, as is typical of conventional constrained matching. In the current algorithm they are represented by competing objectives each having a consistent physical dimension.
9. Applying well-defined Pareto front isolation [3] to the trade-off curve produces ensemble of globally superior solutions in the case of degenerate solutions, as discussed in Appendix B3.
10. The algorithm gives solution options, insight and predictability through the picture of global optimal trade-off, allowing the user to choose what best meets the agenda at hand. This is in contrast to a black-box algorithm giving a single point solution with no insight or context.
11. Likewise, the user can locate the point of diminished return from the trade-off picture to get the best "bang for the buck", not possible in a black-box algorithm. In some cases, such points can be systematically located through analytical formulas.

Each point above is a challenge, if not impossibility, to conventional matching approaches. In a sense, conventional matching carries out considerable computation leading to the single final answer and discards all intermediate results. In the current algorithm, computation is carried out economically with intermediate results kept to form a global picture. These advantages pertain to the optimization algorithm itself, enjoyed by not only matching, but any problem that can be cast in this optimal trade-off format.

## IV. IMPLEMENTING DISTRIBUTED MATCHING

In this section we will explore a matching scheme taking advantage of the algorithm presented above, leading to a deterministic, rigorous and efficient operating paradigm. This scheme consists of the following components:

- Distributed matching through adiabatic reduction of mismatch,
- Interpolation of matching solution from pre-calculated database.

As operation modules these two schemes can be independently implemented, although their combined application promises to deliver maximal gain in efficiency and effectiveness in terms of distributed matching.

### 1. Distributed matching through graduated reduction of mismatch

To implement distributed matching, one first segment the entire beam line into potential matching sections, with all quadrupoles (and skew quadrupoles) in each section being eligible candidates as matching elements. The segmentation scheme is flexible, and does not have to be contiguous. Self-contained transport modules, such as acceleration modules or dispersion suppressors, can be either left out of or embedded inside a matching section. These are shown in Figures 4(a-d).

The continuous spectrum of optimal trade-off solutions from the algorithm described in Section III0 provides the basis for distributed matching, where an initial mismatch $\Phi$, such as defined in equation (3), is adiabatically reduced to unity over many sections, maintaining optimal trade-off between $\Phi$ and K at each intermediate stage. The executor of this algorithm has the freedom to decide how to taper the profile of $\Phi$ reduction over successive sections, based on the trade-off curves made available to him <u>for each section</u>. This concept is depicted in Figure 4(e).

In this construction the matching target for each section is the <u>design</u> beam covariance $\Sigma_D$ of equation (4) at its exit, a <u>static</u> quantity independent of either the incoming beam ($\Sigma_M$) or how the mismatch reduction profile is tapered. This is an important point to keep in mind. By definition $\Phi$ stays unchanged across any section not used for matching.

### 2. Distributed transport error correction

To adiabatically correct an error $\Phi$ as defined in equation (6) originating from a measured local deviation from design transport, as shown in Figure 5(a), the procedure outlined in the previous section can be readily applied, with the measured beam covariance $\Sigma_M$ in equation (3) replaced now by $\Sigma_M^{Out}$ of equation (5) determined by the measured transport error. The equivalent beam mismatch $\Phi$ is now given by equation (6) with *M* being the measured transfer matrix across the transport error. Correction of transport error is accomplished when an on-design beam, $\Phi=1$, at the entrance P of the overall transport (Figure 5) is again on design at the exit Q (Section IIB2). One can repeat the above procedure and bring $\Phi$ adiabatically to unity over successive downstream sections as shown in Figure 5(b). An alternative is to cancel this transport error by "front loading" the matching solution over <u>upstream</u> sections as shown in Figure 5(c). This can be done by the following procedure:

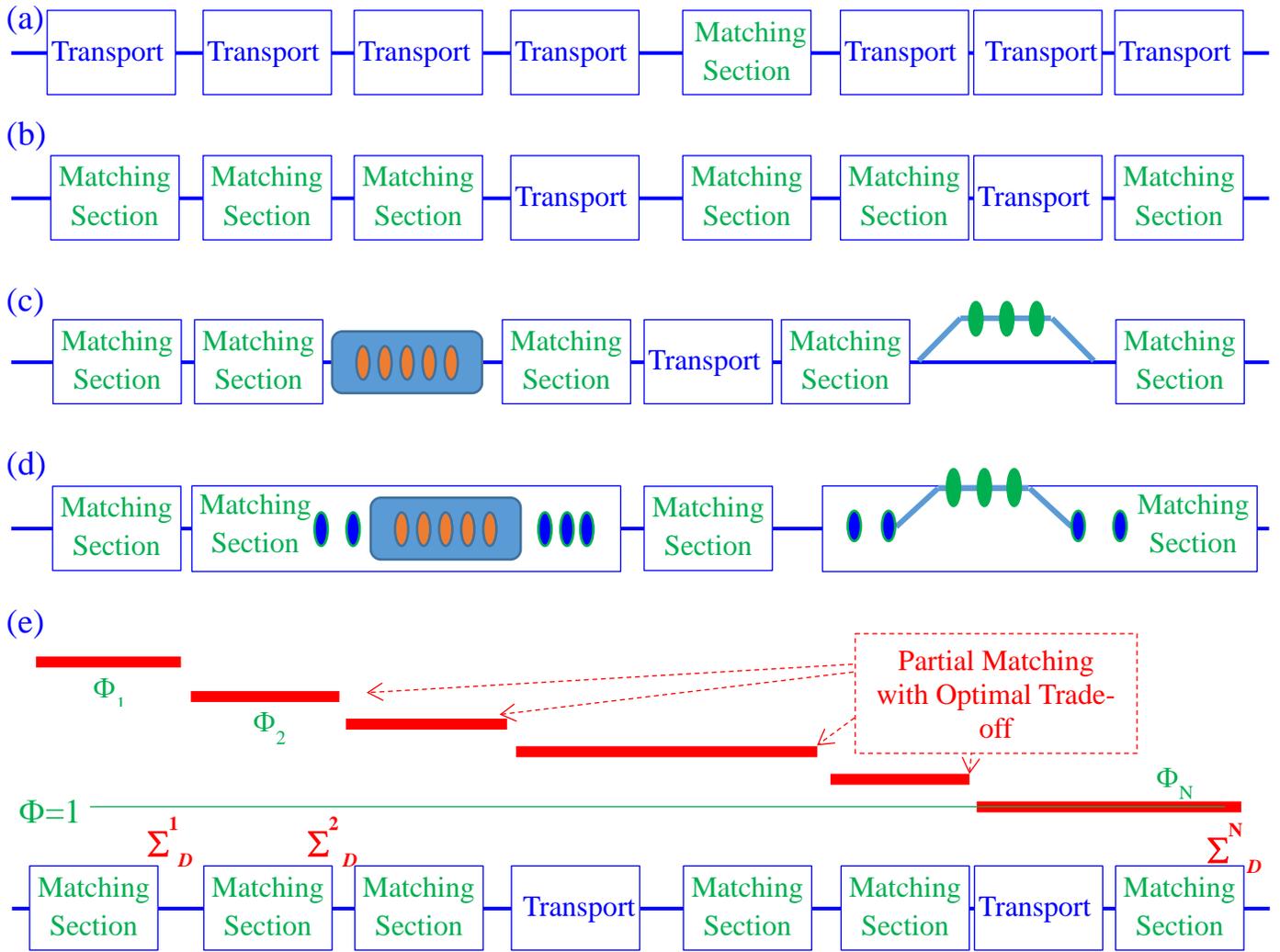

Figure 4. Concept of distributed matching. (a) Localized matching (b) Segmentation into distributed matching sections, (c) Special modules left out, (d) Special modules embedded, (e) Adiabatic reduction of beam mismatch $\Phi$, represented by level of thick red lines, by <u>partially</u> matching to design covariance $\Sigma_D$ at each section until $\Phi=1$. $\Sigma_D$ is static and never changes.

1. Launch <u>reverse</u> <u>design</u> beam covariance $\Sigma_D^{Rev.}$ formed from reverse beam coordinates (See equation (4))

$$\Sigma_D \xrightarrow[\substack{x \to x \\ x' \to -x' \\ \beta \to \beta \\ \alpha \to -\alpha}]{} \Sigma_D^{Rev.} \quad (16)$$

through the <u>inverse</u> of measured transport error matrix $M^{-1}$ to get <u>reverse</u> initial mismatched covariance

$$\Sigma_M^{Rev.} = M^{-1} \cdot \Sigma_D^{Rev.} \cdot M^{-1^T}. \quad (17)$$

2. Geographically invert the entire line preceding the transport error and segment it into distributed matching sections, with any transport optics $M$ between matching quadrupoles replaced by its inverse $M^{-1}$.

3. Use the inverted line to perform distributed matching on $\Sigma_M^{Rev.}$ as before by adiabatically bringing $\Phi$ down to unity across many sections. Now matching target becomes the design covariance $\Sigma_D^{Rev.}$ at the end of each <u>inverted</u> section, which is related to design covariance at the <u>beginning</u> of each original section through the transformation (16).

4. The quadrupole solutions thus obtained will collectively modify the <u>upstream</u> transport necessary to pre-empt the measured transport error.

Instead of inverting the upstream beam line, we could have also kept the original beam line, redefined $\Sigma_M^{Rev.}$ of (17) entering the transport error section as the <u>new</u> <u>design</u> beam covariance, and simply used the algorithm

to match to this new design:

$$\Sigma_M^{Rev.} \xrightarrow[\substack{x \to x \\ x' \to -x' \\ \beta \to \beta \\ \alpha \to -\alpha}]{} \Sigma_D^{New} . \tag{18}$$

The reason for using the inverted line will become clear in the next section in the context of interpolated matching solutions.

Finally, one can choose to further soften the impact on baseline transport by splitting the difference between front loading and back loading over an even wider expanse of the beam line around the error as shown in Figure 5(d).

The possibility to front load transport correction through distributed matching is actually key to expanding the scope into tailoring local optics for purposes beyond matching. The algorithm developed thus far provides a systematic, flexible, and rigorous way to do this. An example will be given in Section IV4.

### 3. Interpolation of matching solution from pre-calculated database

The design matching targets $\Sigma_D$ at the end of each matching section shown in Figure 4(e) do not change, regardless of incoming mismatch, mismatch reduction profile, or changes internal to the matching section such as (knowable) RF focusing. This significantly narrows the scope of any effort to map out the landscape of solutions for a given matching section. One only needs to pay attention to variations in input beam and possibly internal parameter changes, but never the static matching target. An interpolation scheme thus can be conceived in which a comprehensive table of matching solutions, or rather trade-off curves, is pre-computed as function of input beam and internal optical parameters (e.g., RF phase). During operation this table can be used to give immediate and proven matching solutions by interpolation. Such a scheme ensures a matching platform with significant robustness and efficiency, as all solutions are already worked out and proven correct. There is no room for surprise, and cost to beam time by matching computation is negligible.

Such an interpolation scheme has no logical interdependency with distributed matching, but each scheme gains enhanced effectiveness and versatility when combined with the other. In addition, the deterministic algorithm introduced in Section III is ideal as the engine for generating the interpolation table for two reasons. Firstly, it is out of the question to generate such a massive table if case-by-case parameter tweaking is required as is typical of many conventional methods. Secondly, successful interpolation as described strongly depends on smooth dependence of the function to be interpolated on input variables, which cannot happen with guesswork and random number generator incorporated into the process.

Distributed matching gains tremendously in efficiency with such an interpolation framework, as at each stage of $\Phi$-reduction the resulting $\Sigma_M$ can be quickly fed into the interpolation engine to obtain the partial solution for the next stage[2], until $\Phi=1$ is reached. Since the matching target $\Sigma_D$ at the end of each section is static, the interpolation table needs be pre-computed only <u>once</u> per section. In the special case of a periodic lattice where each period is a designated matching section, only one table needs be calculated for <u>all</u> matching sections.

The interpolation table is multi-dimensional, with axes corresponding to two types of independent variables (See Figure 6(b)):

1. Parameters characterizing incoming beam covariance. These can be most logically taken to be the amplitude and orientation of the beam in the phase space normalized by design Twiss parameters,

$$\begin{aligned} x &\to x/\sqrt{\beta_D} , \\ x' &\to (\alpha_D \cdot x + \beta_D \cdot x')/\sqrt{\beta_D} , \end{aligned} \tag{19}$$

in both the X & Y planes. In this space the incoming mismatched beam in each plane is an ellipse uniquely defined by two parameters, length of the semi-major axis $\Lambda$, and its angle relative to the reference axis $\Theta$, as shown in Figure 6(a). $\Lambda$ and $\Theta$ completely determine the incoming mismatched Twiss parameters $\beta_M$ and $\alpha_M$:

---

[2] This can be output by the algorithm. In situations prone to measurement and setting errors, empirically measured mismatch at each stage can be a more reliable input to ensure convergence.

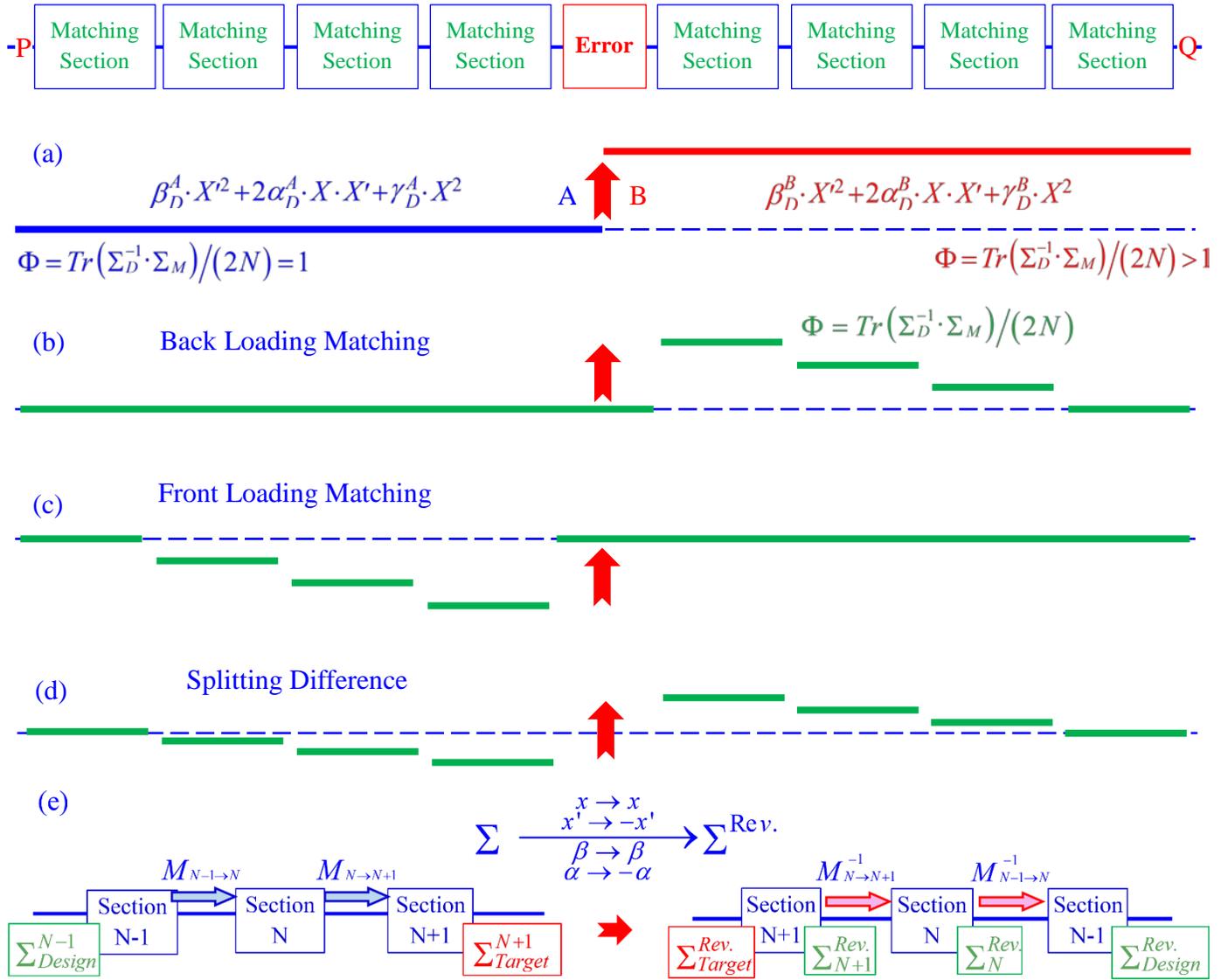

Figure 5. Scenarios to implement transport error correction . (a): The Courant-Snyder invariant *C* of equation (2) executes a jump across the transport error (red arrow). We can recast the transport error as an equivalent beam mismatch Φ>1 of equation (3) for the on-design input beam. The goal of transport error correction is then to bring Φ back to 1. This can be done by back loading the distributed matching solution (b), front loading the solution (c), or splitting the difference (d). (e): Concept of front loading distributed matching in inverted lattice.

In order to adiabatically go from a matched state (Φ=1) to a pre-defined mismatched $\Sigma_{Target}$ without changing the design matching target ad hoc all the time, the lattice is inverted and the roles of initial and target beam covariances, $\Sigma_{Initial}$ and $\Sigma_{Target}$, are switched. The target beam covariance becomes the initial covariance and the goal is again to bring Φ down to one, with target covariances at each inverted section related to design at the entrance of the original sections. Thus again the target covariance is static and only needs be computed once.

$$\beta_M = \frac{\left(\Lambda^4 \cdot \cos(\Theta)^2 + \sin(\Theta)^2\right) \cdot \beta_D}{\Lambda^2},$$

$$\alpha_M = \frac{\left[\begin{array}{c}(\Lambda^4 - 1) \cdot \cos(\Theta) \cdot \sin(\Theta) \\ + \left(\Lambda^4 \cdot \cos(\Theta)^2 + \sin(\Theta)^2\right) \cdot \alpha_D\end{array}\right]}{\Lambda^2}, \quad (20)$$

$$\Lambda^2 = \Phi + \sqrt{\Phi^2 - 1}.$$

Thus any beam with mismatch in both X & Y can be specified by 4 numbers: $\Lambda_X$, $\Lambda_Y$, $\Theta_X$, and $\Theta_Y$. Similar parametrization can be worked out for fully XY-coupled systems involving more parameters, although the interpolation scheme discussed here has much more limited applicability to such systems. A prototype of such table is shown in Appendix B7 (Figure B7), which already can be regarded as containing solutions to all distributed matching problems in a FODO lattice.

2. Parameters characterizing state of special transport modules embedded in the matching section. An example of this can be an RF cavity whose transfer matrix is defined through the RF phase and amplitude (for fixed momentum). These parameters clearly have effect on the matching section transport and need to be included as variables. This is illustrated in Figure 6(b).

The interpolation table of all trade-off curves is calculated by the deterministic algorithm discussed in Section III over the space spanned by mismatch parameters $\Lambda_X$, $\Lambda_Y$, $\Theta_X$, and $\Theta_Y$ (first type), and when applicable, parameters of embedded modules (second type). Again the constant matching target $\Sigma_D$ at the end of each section means this only has to be done once. In real operation, when a beam mismatch (Section IV1) or transport error (Section IV2) is measured, one can quickly look up the pre-calculated table for corresponding mismatch and embedded module parameters and interpolate for necessary trade-off solution. This process is passed from one section to the next until the design matching target is reached. Figure 7 gives an illustration of this iterative process.

It is important to note that, given measurement and implementation errors, it is preferable to empirically measure the outcome of each matching section before it is fed into the next section, rather than relying on theoretical prediction. This should help accelerate the convergence of the process. For both beam profile matching and transport error correction this empirical data to calibrate input to interpolation at every stage can be achieved through local transfer matrix measurements. Fortunately this is a well-defined operational procedure in accelerators (e.g., see [1]).

Once the interpolation table is created, online matching efficiency is limited only by that of interpolation, a well-controlled process. Time and resource requirements for matching computation does not enter operational consideration. All optical manipulation computations can be acquired and implemented within a minimal and predictable time frame.

The front loading scenario for transport error correction discussed in Section IV2 poses a particular challenge in the context of interpolation and deserves some elaboration, which explains the need for an inverted scheme. The reason we only need to calculate one set of interpolation tables good for any input situation is because the matching target $\Sigma_D$ is the static design covariance at the end of each section. If we were to execute a front loading solution by adiabatically increasing $\Phi$ from the front end of the line at $\Phi=1$ to the entrance of the optical error where $\Phi$ is increased to the proper level to cancel the error, we would need interpolation tables with not only variable initial mismatch, but also variable target mismatch or, in other words, an infinite number of interpolation tables. In addition, there is no clear logical guideline as to what to set as the matching target at the end of every intermediate section. By inverting the lattice and taking as matching target $\Sigma_D$ at the beginning of each section, the target is again a unique and static quantity and only one table is needed per inverted section. Figure 5(e) describes this process.

The front loading scenario also has special implication for using empirically determined mismatch from one section as input to interpolation for the next section. In this scenario the partial matching solution should be loaded into the machine in reverse order, first into the last section right before the optical error, and then worked backwards toward the first section. At the end of each stage the transfer matrix $M_N$ of the current matching section $N$ should be empirically measured and used to back-propagate the previous matching outcome $\Sigma_M^N$ at the entrance to the downstream section $N+1$ to get

$$\Sigma_M^{N-1} = M_N^{-1} \cdot \Sigma_M^N \cdot M_N^{-1^T} , \qquad (21)$$

where $\Sigma_M^{N-1}$ is the reversed covariance (equation (16)) at the end of the upstream section $N-1$. $\Sigma_M^{N-1}$ is then used as interpolation input for matching the next (upstream) section $N-1$. This reversal of solution loading order is for exactly the same reason as above. Had the solutions been loaded starting from the most upstream section, any discrepancy in resulting measured transport would enforce a change in matching target in the next section, and thereby require an infinite number of interpolation tables to chase after a moving target.

Some points should be noted regarding practicalities of creating the interpolation table.

- Generating interpolation table on a massive scale may benefit from using thin lens quadrupoles in the integration for efficiency. One can use a recipe developed in [2] to convert from thin to thick lens solution with very high fidelity. This conversion can be iterated for higher precision, although the iteration process itself cannot be easily put on an interpolation framework.
- When the procedure of Pareto front isolation (Appendix B3) is applied, a discontinuity can be introduced to the paths traced out by all quadrupoles, $k_1$, $k_2$, …. $k_N$, along the trade-off curve. While this poses no problem to executing intermediate matching in individual mismatched cases, when combined into an interpolation table the misalignment in the discontinuities between different curves will cause problem. For example, the trade-off curves corresponding to adjacent values of mismatch amplitudes, $\Lambda_1$ and $\Lambda_2$, have slightly different discontinuities at $\Phi=2.0$ and $\Phi=1.8$ respectively. Now if the input mismatch amplitude is half way between $\Lambda_1$ and $\Lambda_2$, and the user wants to select an intermediate matching solution at $\Phi=1.9$, he will end up interpolating two values belonging to <u>different</u> branches of the trade-off curve from $\Lambda_1$ and $\Lambda_2$. The way to circumvent this is to keep track of such discontinuities and artificially force the interpolation inputs to be always on the same side of the discontinuity in $\Phi$. The resulting offset is negligible if the table samples the discontinuity region with high enough resolution. Again in the above example if the sampling near the discontinuity is so dense in $\Lambda$ such that the discontinuity occurs at $\Phi=1.79999$ and $\Phi=1.80000$ in adjacent curves for $\Lambda_1$ and $\Lambda_2$, the interpolation can then simply be forced to always stay outside this negligible gap with no practical consequence.

As mentioned earlier, the interpolation scheme has no logical interdependency with distributed matching, but each scheme is more effective when combined with the other. This is because together the two schemes enable a degree of error tolerance through iteration. If an error is introduced during one step of interpolation, it can be captured in the next segment of distributed matching, whereas any residual mismatch left from the previous distributed matching segment becomes input to the interpolation for the next segment. Possible inefficiency caused by the need to perform distributed matching is more than compensated by the very efficient interpolation process bypassing online computation. In an even more computationally efficient variation (requiring more extended distributed matching sections) one can even do without interpolation and directly apply the solution in the interpolation table closest to the incoming mismatch every time.

### 4. Jitter suppression by optical re-matching

In this section we give an example of using distributed matching to systematically suppress beam orbit jitter by reshaping the design optics. The procedure and algorithm developed so far for distributed matching can easily achieve this goal. We will focus on orbit jitter coming from localized sources, therefore occupying a relatively correlated slice of the phase space, as opposed to random jitter homogeneously filling up the phase space making suppression difficult even for dedicated feedback systems.

For simplicity we consider an orbit jitter caused by a single source such as a defective horizontal dipole power supply. This launches an orbit jitter signature $(x_J, x'_J)$ everywhere downstream. The Courant-Snyder invariant $C$ of equation (2) for this signature, measuring the invariant "action" <u>relative to design optics</u>, is simply

$$C = \frac{x_J^2}{\beta_D} + \beta_D \left( x'_J + x_J \frac{\alpha_D}{\beta_D} \right)^2 . \tag{22}$$

Clearly minimum of $C$ occurs when

$$\begin{cases} \alpha_D = -\beta_D \frac{x'_J}{x_J} \Rightarrow C \to 0 . \\ \beta_D \to \infty \end{cases} \tag{23}$$

In practice $\beta_D$ cannot be too large of course, but equation (23) provides guidance for a matching target to aim for. By "shaping" a new <u>design</u> optics conforming to (23) the action of the jitter can be de-magnified at a specific location, after which the <u>design</u> optics can be re-matched to the original design while the action $C$ stays small. A conventional dedicated matching section cannot perform this function since this needs to be carried out at arbitrary locations over extended areas. This procedure is depicted in Figure 8.

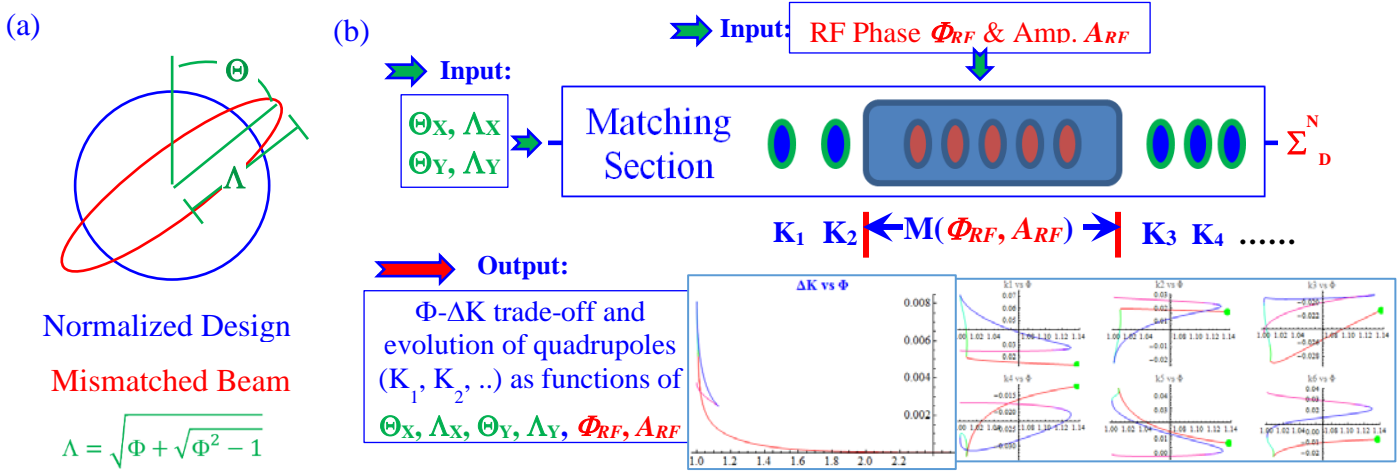

Figure 6. (a): Representing single-plane mismatch by beam ellipse parameters $\Lambda$ and $\Theta$ in the normalized space. (b): Concept of pre-calculation of interpolation tables to be used for online distributed matching. The table is spanned by independent variables of two types marked as input in (b): mismatch parameters $\Lambda$ and $\Theta$ in both planes, and any operation parameters in the embedded modules contributing to transport, such as RF phase. Once the table is generated, straightforward interpolation based on these two inputs will give the corresponding trade-off curve and quadrupole solutions, as those shown in Appendix B (example plots taken from Figures B3 & B4).

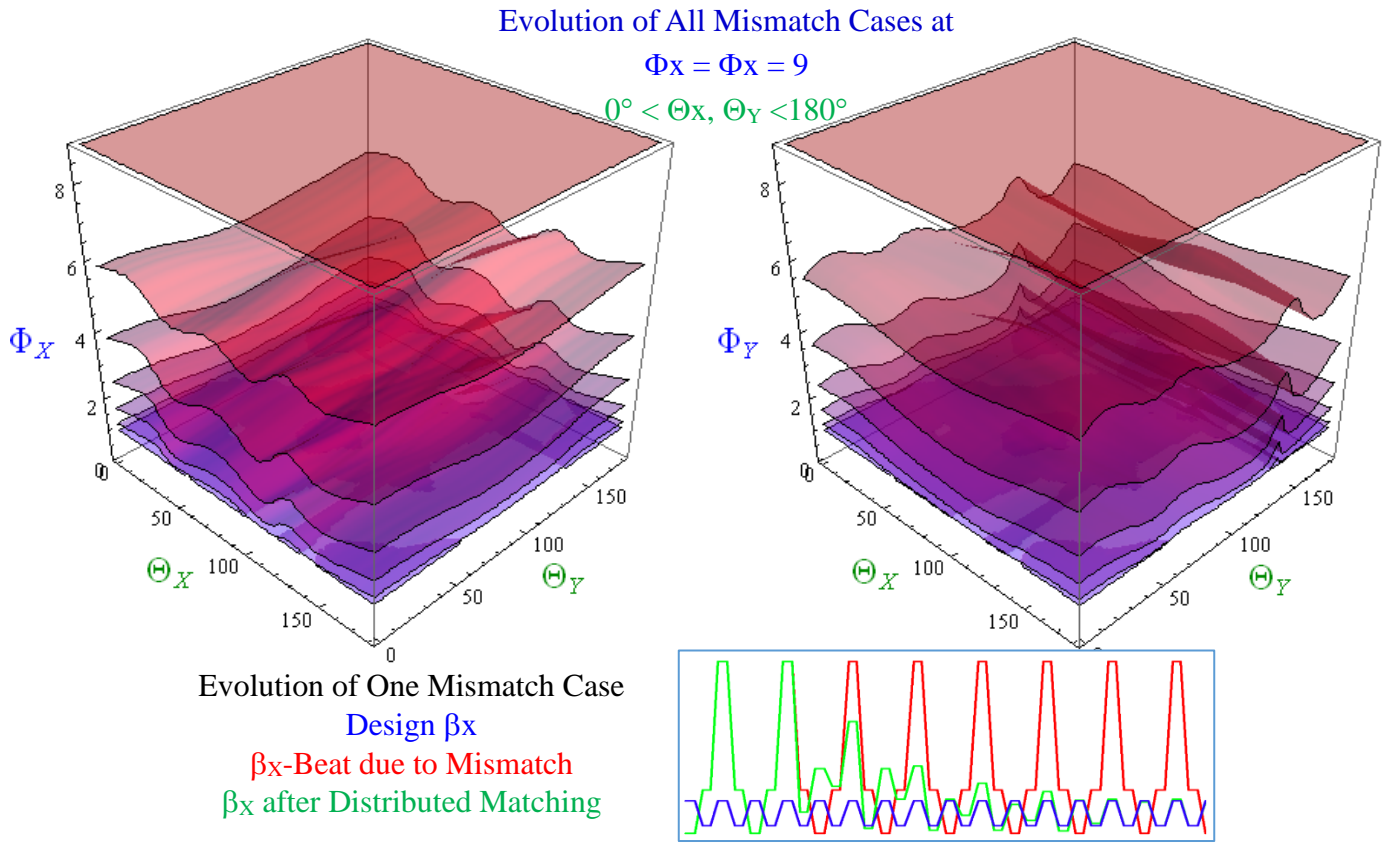

Figure 7. Distributed matching by interpolated solution. Evolution of initial mismatch $\Phi_X=\Phi_Y=9$ ($\Lambda=4.236$), and $\Theta_X$ and $\Theta_Y$ covering the entire range 0–180° (top sheet in red), over multiple partial matching sections (subsequent sheets from red to blue corresponding to decreasing $\Phi$), in a 120° FODO lattice. Each section consists of 3 quadrupoles of alternating polarity. At each stage and for each point on the sheet the partial matching solution for the next stage is <u>interpolated</u> from a <u>pre-calculated</u> table of partial matching solutions. The evolution of a single point from $\Phi=9$ to $\Phi=1$ is exemplified in the bottom graph showing design $\beta$ (blue line), mismatched $\beta$ (red), and evolution of $\beta$ by distributed matching (green). The entire set of initial $\Phi=9$ for all possible $\Theta$'s is brought down to $\Phi=1$ everywhere (bottom blue sheet) within 7 iterations by interpolated solution alone.

$C$ of equation (22) can be recast in a form akin to $\Phi$

$$C = Tr(\Sigma_D^{-1} \cdot \Sigma_T), \quad \Sigma_T = X \cdot X^T, \quad X = \begin{pmatrix} x_J \\ x'_J \end{pmatrix}, \quad (24)$$

from which one can estimate the de-magnification of $C$ in the procedure outlined in Figure 8,

$$R = \frac{Tr(\Sigma_D^{-1} \cdot \Sigma_T)}{Tr(\Sigma_O^{-1} \cdot \Sigma_T)}$$

$$\xrightarrow{\alpha_o = -\beta_o \frac{x'_J}{x_J}} R = Tr(\Sigma_D^{-1} \cdot \Sigma_O) + \frac{\beta_D}{\beta_O} = \Phi_{Beam} + \frac{\beta_D}{\beta_O}$$

$$\Sigma_D = \begin{pmatrix} \beta_D & -\alpha_D \\ -\alpha_D & \gamma_D \end{pmatrix}, \quad \Sigma_O = \begin{pmatrix} \beta_O & -\alpha_O \\ -\alpha_O & \gamma_O \end{pmatrix},$$

$$\alpha_O = -\beta_O \frac{x'_J}{x_J},$$

(25)

where subscript $D$ is for the original optics and $O$ is for the new optics created at the jitter source to satisfy (23) with a reasonably large $\beta$. It is clear from (25) that this procedure causes intermediate β-beat ($\Phi_{Beam}$) commensurate in size with jitter de-magnification.

A realistic example of applying this to a 30° FODO lattice is shown in Figure 9. Front loading and back loading schemes of Figure 5 are applied to distributed matching sections 1 and 2 respectively, each consisting of 18 quadrupoles, on both ends of the jitter source. More steep optics change can be used to achieve same purpose using less quadrupoles. If pre-calculated interpolation tables exist for both forward and inverted lattices, and proper loading order is followed as described in Section IV3, this procedure can be dynamically implemented in a beam line with no online computation needed other than that for interpolation.

An important point to keep in mind, which is the entire point of optimal trade-off, is that the above is accomplished with minimal disturbance (ΔK) to existing optics. There may be other ways to achieve the same jitter suppression, but this algorithm guarantees the least invasive configuration change.

This example also testifies to the robustness of the integration algorithm of Section III, as a 30° lattice does not lend itself well to matching due to poorly decoupled $\beta_{X/Y}$. A less robust algorithm would find it difficult to accomplish this task.

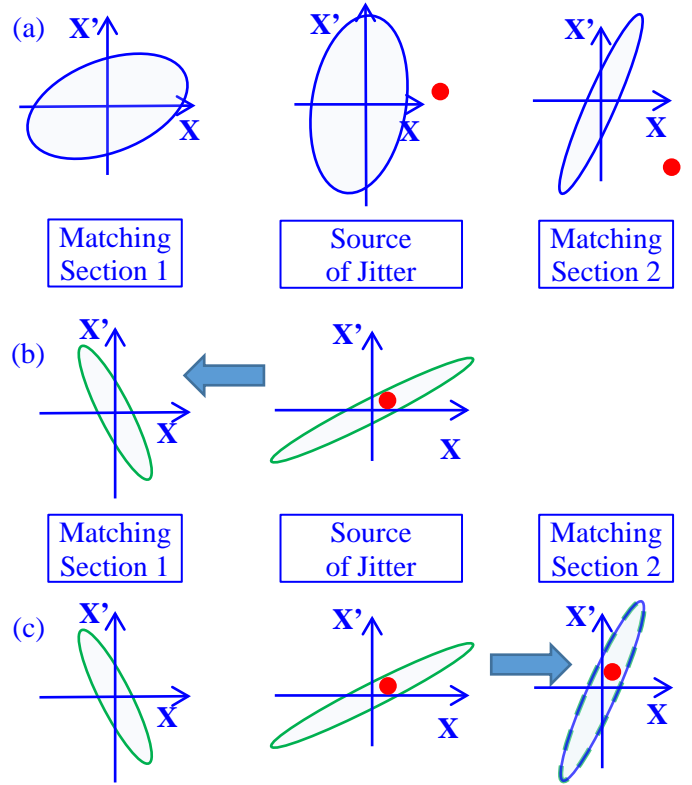

Figure 8. Suppressing orbit jitter with optics manipulation by distributed matching. (a): Jitter displays phase space signature indicated by red dot after the source, and at downstream point. (b): Design beam phase space is redefined to maximally conform to jitter after the source, and beam is matched to new design by distributed matching. (c): Design beam is restored to original design downstream by distributed matching. Jitter amplitude is reduced.

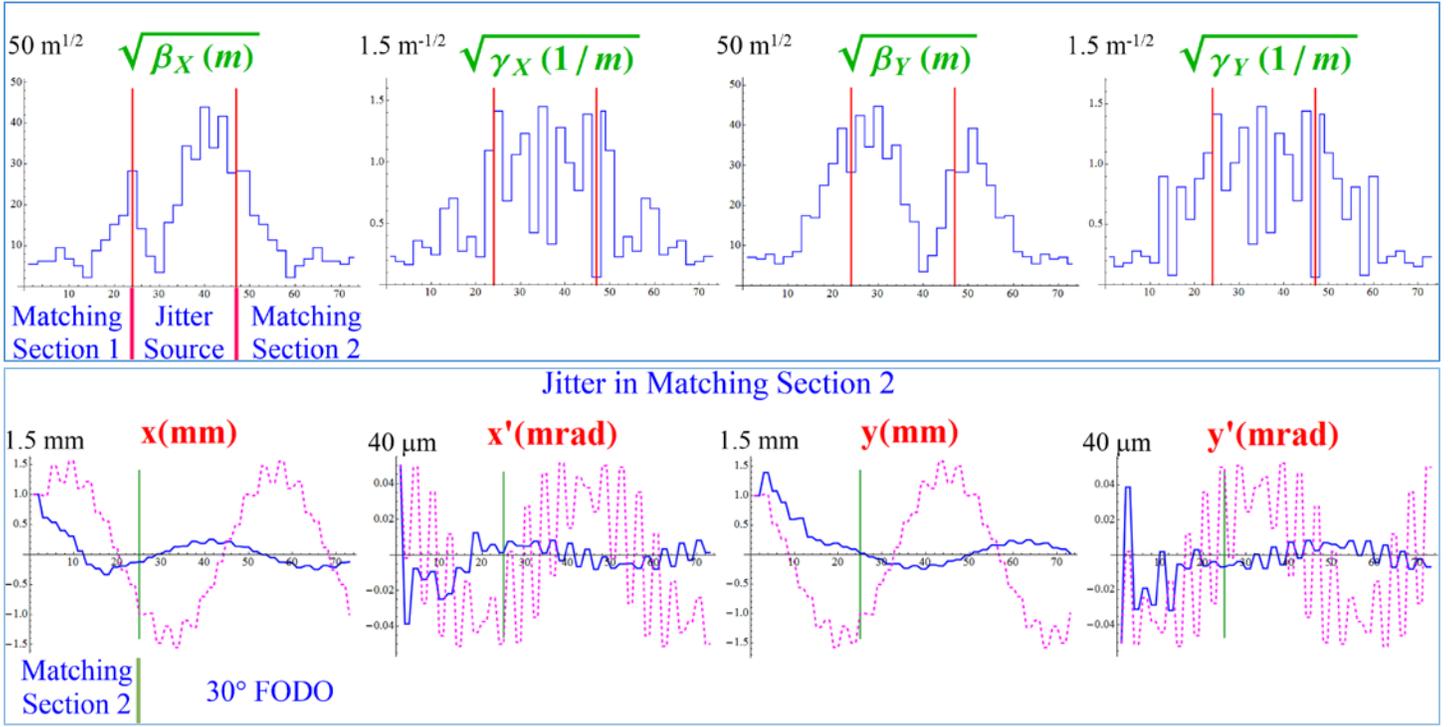

Figure 9.  Suppressing orbit jitter with optics manipulation by distributed matching – Realistic example in a 30° FODO lattice.   Top: Twiss parameters of first matching section (to match beam to new design), region of jitter, and second matching section (to restore beam to original design).  Bottom: Evolution of jitter inside second matching section followed by downstream FODO section, with a reduction by roughly one order of magnitude.  Magenta dashed (blue solid) line shows unsuppressed (suppressed) jitter.  A $\beta_D$ of 800 m is used (see equation (23)).  Matching based on a 30° lattice is fundamentally more challenging than stronger focusing lattice, but presented no problem to the deterministic algorithm introduced in Section III.

## V.  EXTENSION AND OTHER APPLICATIONS

Freeing up most quadrupoles in a beam line not only provides a more flexible and efficient way to perform matching, less plagued by problems associated with localized matching, but also enables optics manipulation over extended range for purposes beyond matching, as described in the last example.  Scheme described in Section IV3 can further allow such manipulations "on the fly" in a fully predictable way.  The integration algorithm developed in Section III and Appendix B provides a rigorous and deterministic basis for this manipulation.  It should be clear that this algorithm is not limited to matching either, but works for any optimal trade-off between well-defined objective functions $F$ and $H$, which can be performance metrics of the beam or a tuning process, or hardware specs.  It should also be noted that weighting can be applied inside the objective function, with physically consistent meaning.  For example instead of (7) one can have

$$H = \mathrm{K}(k) = \sum_{m=1}^{N_Q} w_m \cdot k_m^2 ,  \qquad (26)$$

where $w_m$ is a weight on the quadrupole strengths to favor changing some quadruples over the others.  This weighting is conceptually unambiguous, as opposed to weighting between non-congruent quantities, such as $\beta$-function and quadrupole strength, often found in constrained matching.

It is also conceivable to use as independent variable the common focusing strength shared by quadrupoles on a power supply string if the entire string does not have too long a span.  Such a scheme would not be easily accessible for conventional matching algorithms, especially when the only option is to achieve complete matching.  On the other hand in the algorithm presented here the common focusing strength, or any additional focusing through trim power supplies, can enter the formulation as variables and optimally partial-matched

### 1. Matching in XY-coupled systems

Nothing in the trade-off algorithm developed so far precludes its application to matching in a fully XY-coupled system. The mismatch factor $\Phi$ of (3) or (6) accounts for off-diagonal elements in the covariances. Appendix A applies equally well to XY-coupled cases, with $\Phi=1$ uniquely qualifying a fully 4D matched beam. One only needs to include skew-quadrupoles with enough degrees of freedom to achieve this goal. In case of insufficient degrees of freedom the algorithm will stop ($\lambda=0$) at the point with lowest $\Phi$ possible within the monotonic optimal tradeoff regime. It is useful to note that as all 4 diagonal components of the quantity $\Sigma_D^{-1} \cdot \Sigma_M$ in (3) are positive-definite, we will not encounter run-away situations where large values are generated to fine-cancel to a small one.

A special variation of the XY-coupled matching involves rotating quadrupole channels [4] where the addition of a rotation degree of freedom to normal quadrupoles eliminates the need for dedicated skew quadrupoles and allows coupling manipulation in a more distributed manner. The current algorithm is readily applicable to such cases by including the rotation angles as additional arguments to the transfer matrix $M$ in equation (6).

In addition to using the algorithm to simultaneously perform matching and eliminate XY coupling in initially coupled beam or transport, the ability to manipulate fully coupled 4D phase space in a <u>distributed</u> scheme delivers the same advantage over localized schemes as discussed in Section IIA and Figure 1. The following are two such possibilities:

- Emittance allocation between X & Y

The ratio between X & Y beam emittances has particular significance in many accelerators, especially colliders [5]. This ratio is determined by the 4D beam covariance $\Sigma_D$. Thus with suitably defined <u>design</u> values of $\Sigma_D$ at all locations corresponding to desired XY emittance allocation as in Figure 4, the same recipe developed so far is readily applicable to bring the beam to final matched state with desired X & Y emittance allocation.

- Transport in terrain-conforming beam lines

When the accelerator stops being planar due to topographical constraints [6], nontrivially XY-coupled transport becomes a reality. While at design level this can be kept under control, misalignment and field errors can complicate real beam transport with severe consequence on beam quality. In such cases accurate means of determining the full 4D transport at all locations is of first order importance. Given this input, a transport correction scheme using all focusing elements distributed over the entire line as suggested in Figures 4 and 5 can again deliver all the advantages not possible with a local matching configuration.

Details and potential caveats of implementing the above schemes are beyond the scope of this report. Nevertheless a distributed scheme enabled by the current algorithm allows graduated manipulation of XY-coupled beam/transport by generic elements, Especially if rotating quadrupoles are an option, throughout the line, with multiple advantages over dedicated, localized schemes. The trade-off integration algorithm furthermore provides a systematic, deterministic and rigorous recipe to realize this.

### 2. Trade-off between mismatch factors $\Phi$

In this section we discuss applications of the trade-off algorithm when both objectives, $F$ and $H$, are mismatch factors with respect to different "design" optics. This usually happens when there are competing goals for beam phase space or machine transport. The algorithm presented in this report can produce a continuous "knob" through which the user can systematically survey the landscape of trade-off between $F$ and $H$, and decide on the best compromise working point(s) in between. Every point thus selected represents the optimal trade-off between these two matching options at different level of trade-off.

- Trade-off between beam and orbit jitter

One can encounter situations where phase space characteristics of undesirable orbit oscillation or jitter is strongly incongruent with design <u>beam</u> profile. In other words, the Courant-Snyder invariant, or action $C$ of equation (24), of the jitter with respect to design optics at a fixed amplitude is near maximum compared to other design optics alternatives. Adjusting optics to accommodate this jitter would maximally compromise beam matching and vice versa. Carrying out the jitter suppression program outlined in of Section IV4 may be either too elaborate, or simply impossible because the jitter source is in the injector, such as the helicity-correlated orbits that can compromise measurement precision in parity violation experiments in nuclear physics [7]. In this case one can establish the continuous trade-off curve connecting the two "design" optics matched to the beam and the oscillation respectively, from which the best compromise optics can be continuously sampled and decided on to minimize their collective impact.

- Trade-off between beam and halo

Beam halo can pose serious limitations to accelerator performance, especially in high current operations. Apart from dedicated collimation schemes, online control can be an additional measure to further limit its impact by matching the halo phase space profile to machine acceptance. This profile can be incongruent with design beam covariance, and application of the trade-off algorithm can again help sample and identify the best compromise operating point(s).

- Trade-off on emittance allocation between X & Y

X-Y emittance allocation was discussed in the previous section in the context of 4D matching via a distributed configuration. One can also envision trade-off between different emittance allocation scenarios through beam covariances $\Sigma_1$ and $\Sigma_2$. The algorithm is then applied with competing objectives being 4D mismatch factors $\Phi_1$ and $\Phi_2$ with respect to $\Sigma_1$ and $\Sigma_2$.

- Trade-off between longitudinal slices in FEL driver beam to achieve fresh-slice lasing

In self-amplified (SASE) FEL sustained lasing by the same electron bunch results in degraded energy spread, and in turn reduced lasing efficiency and quality. It is useful to have a scheme where lasing sites lengthwise within an electron bunch can be "switched on" by turns so that there is a continuous supply of "fresh" lasing slices. If this scheme is embedded inside the undulator channel, one can produce quality laser pulses with different characteristics, each one from a fresh slice in the same electron bunch. A "2-color" and "3-color" fresh slice scheme, employing differential orbit offsets across longitudinal slices to maintain slice freshness, has been demonstrated [8]. It is conceivable that one can also introduce differential mismatch across longitudinal slices for the same purpose, and the current algorithm can produce a knob to move the lasing site from the head to the tail continuously inside the undulator. Timing between light pulses emitted from different slices can be controlled by magnetic chicanes [8]. The differential mismatch can first be introduced into the beam via one of the following mechanisms:

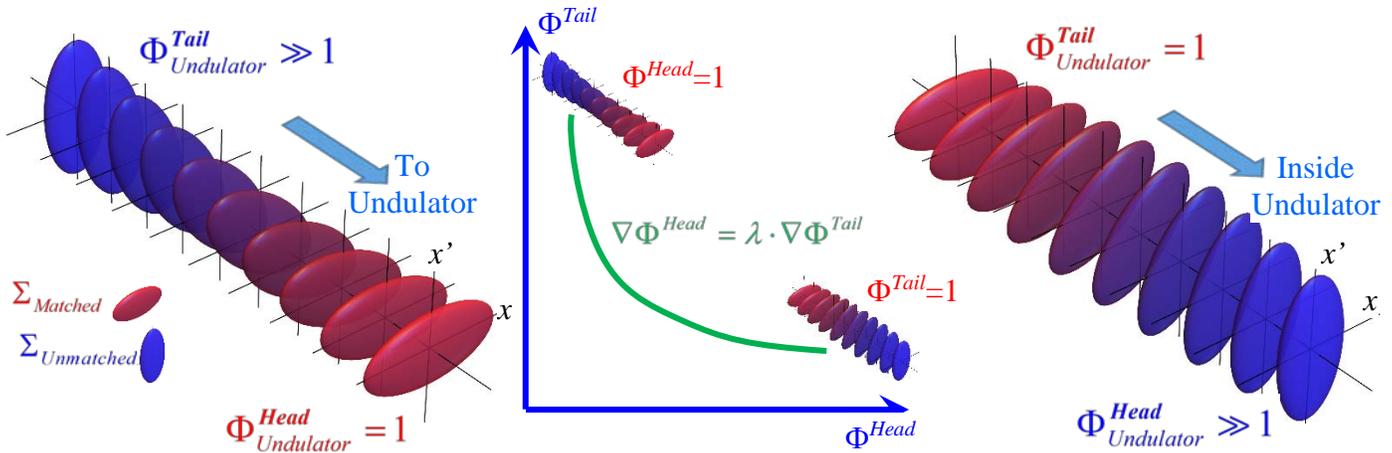

Figure 10. Concept of slice-dependent mismatch introduced into FEL driver beam and lasing site selection by trade-off on matching targets. Different beam slices along the longitudinal dimension are made to develop different mismatch $\Phi$ with respect to the design beam covariance required for optimal lasing in the undulator. The trade-off algorithm implemented on quadrupoles inside the undulator can be used to continuously switch on successive slices for lasing, while simultaneously suppress other slices to keep them fresh (minimal momentum dilution). Only slices closely matched to the undulator will be lasing and incurring momentum spread at any given matching setting. The rest of the beam executes large amplitude β-beat without lasing. The competing objectives $\Phi^{Head}$ and $\Phi^{Tail}$ are as given in (27).

- Differential space charge focusing along the longitudinal dimension near injection energy of the machine, using similar concept as the emittance compensation technique [9],
- Quadrupole wakefield effects, such as the differential head-tail focusing observed at the dechirper in LCLS at SLAC [10,11],
- Combining deliberate energy chirp in the beam with machine chromaticity to produce the same effect. A dipole version of this technique for controlling emittance growth in linacs has been well documented [12]. Further lasing suppression and reactivation of dormant slices can be achieved by introducing skew focusing in the optics to effect full 4D mismatch and rematch to the undulator. This should prove a well-defined task for the currently proposed algorithm.

This concept and experimental proof-of-principle have been proposed by the author at LCLS, for which the current trade-off algorithm will play a key role in two aspects:

- Providing a continuous knob for selecting fresh lasing sites between the head and tail, each represented by a competing objective in the form of mismatch $\Phi$, which is a variation of equation (6).

$$\Phi_{Head} = Tr\left(\Sigma_D^{Undul.} \cdot M(k_m) \cdot \Sigma_{Head}^{In} \cdot M^T(k_m)\right)/(2N),$$
$$\Phi_{Tail} = Tr\left(\Sigma_D^{Undul.} \cdot M(k_m) \cdot \Sigma_{Tail}^{In} \cdot M^T(k_m)\right)/(2N),$$
(27)

with $\Sigma_D^{Undul.}$ the matched beam covariance into the undulator. The trade-off curve in the space of $k_m$'s links the solutions matching head and tail via a continuous optimal path. The process remains deterministic even without a priori knowledge of either optima of (27), provided the recipe for restoring determinism through artificial constraints is used as described in Appendix B8.

- It is as important to suppress lasing in the idle and spent slices through mismatch, as it is to ensure good matching for the lasing slices. In other words, a large enough mismatch differential must be present between the head and the tail for example. It has been shown in preliminary simulation [13] that in the hard X-ray regime at LCLS a mismatch of $\Phi \geq 3$ is sufficient to suppress lasing over reasonable distance without incurring unacceptable energy spread. In the event that the above mismatch-inducing schemes fail to produce enough mismatch differential, it is incumbent on the trade-off algorithm itself to enhance this differential. Two scenarios based on the trade-off algorithm are expected to deliver on this goal:
  - Forcing the trade-off integration to go beyond $\lambda=0$, as illustrated in Figure B2. Subsequent points of $\Phi_{Head}=1$ for the head contain worse matching cases for the tail, and the mismatch differential can be magnified. In practice the head remains matched while the tail can become progressively more mismatched[3], or vice versa. By allowing $\Phi_{Head} \approx 1$, one can obtain even larger $\Phi_{Tail}$, because beyond the monotonic trade-off regime, $\lambda>0$, and simultaneously degraded matching on head and tail is allowed.
  - Large matching differential artificially obtained through known constraints. In this scenario a quadrupole state leading to severe mismatch and lasing efficiency degradation for the head is pre-determined, which is not difficult. The competing objective $\Delta K$ referenced to this state is then traded off against the tail matching target $\Phi=1$. The resulting solution will be the matching solution for the tail closest to the known severe mismatch for the head. The same procedure is then switched between head and tail.

Using more than 4 quadrupoles greatly enhances the effectiveness of the algorithm, especially if distributed matching scheme is invoked. This however puts considerable demand on the real estate inside the undulator channel in the absence of an embedded multi-quadrupole matching section inside.

Depending on the accuracy in beam phase space and transport property determination, this algorithm can potentially be used to control lasing slices to very high accuracy and determinism, generating multiple fresh slices for high efficiency lasing from the same electron bunch. Furthermore, with conceivable optimization of head-tail matching, undulator optics and undulator hardware design aimed at such purposes, this procedure can potentially contribute toward ultra-short pulse modes of laser operation.

Simulation work is currently under way to optimize experimental parameters and make predictions on the performance. Implementation of the deterministic matching algorithm for this purpose is in progress [14].

---

[3] Mismatched in terms of detailed Twiss parameters instead of $\Phi_{Tail}$, which is fixed in value when the head is matched.

## 3. Multiple Twiss matching targets

The matching algorithm can be extended to cases with more than one set of matching targets, namely with target Twiss parameters at more than one location. This is necessary in the following examples:

- Beam phase space requirements at intermediate points in addition to matching condition at the match point. For example, a waist may be desired at a symmetry point before the final match.
- Simultaneous matching of multiple pass beam controlled by common set of quadrupoles.

It is important to note that in forming the above matching configurations the independent degrees of freedom exerted by the quadrupoles must be adequate for the task in hand. In an uncoupled system, the total degree of freedom per plane actuated by an ensemble of quadrupoles is at most 3, regardless of how many are in the ensemble. If complete satisfaction of more than 3 independent Twiss parameters per plane is required, obviously at different locations, the quadrupole configuration must be such that there are adequate number of quadrupoles to independently address each additional degree of freedom.

In the case of more than one set of Twiss target, one can generalize the matching formulation (6) by extending the dimension of $\Sigma$ and $M$ (Consult (4)) and defining the combined mismatch factor $\Phi$:

$$\Sigma^{In} = \begin{pmatrix} \Sigma_A^{In} & 0 & 0 \\ 0 & \Sigma_B^{In} & 0 \\ 0 & 0 & \ddots \end{pmatrix}, \Sigma^{Out} = \begin{pmatrix} \Sigma_P^{Out} & 0 & 0 \\ 0 & \Sigma_Q^{Out} & 0 \\ 0 & 0 & \ddots \end{pmatrix},$$

$$M(k_m) = \begin{pmatrix} M_{A \to P} & 0 & 0 \\ 0 & M_{B \to Q} & 0 \\ 0 & 0 & \ddots \end{pmatrix},$$

$$\Phi = Tr\left(\Sigma^{Out^{-1}} \cdot M(k_m) \cdot \Sigma^{In} \cdot M^T(k_m)\right) / (2N_{Out}),$$

(28)

where A, B, etc. denote locations of initial Twiss parameters, which can belong to different passes of a recirculating line, and P, Q etc. denote locations of matching targets. Note $\Sigma^{In}$ and $\Sigma^{Out}$ need not have the same dimension and $M$ needs not be square. For example one can have a matching problem where the initial 4×4 $\Sigma_A^{In}$ is matched into a 4×4 $\Sigma_P^{Out}$ at point P, with an additional constraint defined by a 2×2 matching condition in x-plane, $\Sigma_Q^{Out}$, at location Q. In this case the overall $\Sigma^{in}$ is 4×4, $\Sigma^{Out}$ is 6×6, and $M$ is 6×4. Also note (28) is normalized by the dimensionality of $\Sigma^{Out}$, namely, $2N_{Out}$. The effect of equation (28) is to formally add up all mismatch factors from multiple matching requirements to solve for the optimal global solution. It is important to make sure that $\Sigma_A^{In}$ and $\Sigma_B^{In}$ etc. in (28) are unaffected by the matching process itself. In other words, the matching action will not affect any of the initial Twiss parameters, such as may happen in a multiple-pass configuration where matching in the first pass starting from $\Sigma_A^{In}$ will propagate to $\Sigma_B^{In}$ in the second pass if intermediate feedback mechanism is missing. This is similar to the requirements for multiple-pass orbit steering.

For the case where there is not enough quadrupole degrees of freedom to satisfy multiple Twiss parameter targets (greater than 3 per plane). the deterministic algorithm proposed here is even more ideal for unequivocally arriving at the best possible compromise solution with a clear physical significance. When λ=0, the integration process stops and the best tradeoff among Twiss matching targets at multiple locations has been reached, as indicated by the optimal combined mismatch factor $\Phi$.

This formulation displays the closest resemblance to orbit correction in its most general form. We have defined a computation formulation where one can locally demand the Twiss parameters to match user-defined target values at multiple locations by controlling an ensemble of quadrupoles, just like demanding beam positions to match user-defined target values at multiple locations by controlling an ensemble of correctors in orbit correction. The only difference is the nonlinear nature of the (unambiguously defined) formulation and thus possible demand on numerical and computational precision.

Finally we note that weighting can be applied among various $\Sigma_P^{Out}$ in (28) to reflect matching priority, much the same as in orbit correction.

## 4. Application in accelerator design – Other control variables

So far the deterministic algorithm has only been discussed in the context of solving matching problems, namely, changing quadrupole or skew-quadrupole strengths to minimize the mismatch factor. The formulation (6) can be applied outside this context if we consider other dependencies of *M*, such as geometry or orbit-dependent optics:

$$M = M(k_m, L_i, O_j \ldots), \quad (29)$$

where $L_i$ are physical dimensions having impact on optics, such as inter-quadrupole distance, and $O_j$ orbit offset inside baseline nonlinear fields, etc. Such formulation can be a valuable tool for designing accelerators, such as a final focus system. $L_i$, $O_j$ and other such variables can be treated exactly the same way $k_m$'s are treated in the formulation. The only point to note is that the competing objective *H* now takes on the form (Consult (8))

$$H = \Delta \mathrm{K}(\mathbf{k}, \mathbf{L}, \mathbf{O}, \ldots) =$$
$$\sum_{m=1}^{N_Q} (k_m - k_m^D)^2 + W_L \times \sum_{i=1}^{N_L} (L_i - L_i^D)^2 \quad (30)$$
$$+ W_O \times \sum_{j=1}^{N_O} (O_j - O_j^D)^2 + \ldots,$$

where *H* is the quadratic sum of deviation of all above variables from user defined initial values. Weighting factors are applied to variables of different flavors as needed in (30). As note earlier, this weighting has a clear physical meaning, as opposed to traditional matching algorithm where Twiss parameters, quadrupole strengths and other parameters are indiscriminately mixed within the merit function as generic variables.

## VI. SUMMARY

In this report we propose an alternative paradigm for transverse phase space and transport control, in which all quadrupoles (and skew quadrupoles as applicable) are used to bring about more gentle and robust matching or phase space manipulation. This scheme can also actively mitigate adverse consequences caused by measurement and setting errors or algorithm failure in the matching process, as it counteracts such errors on the spot as they arise, as opposed to passively accepting these errors with no recourse. The overall effect is globally more on-design and contained beam and transport, less drastic demands on beam condition and hardware, reduced beam quality degradation from large amplitude aberration and chromatic effects, and more systematic and efficient operational procedures.

The distributed scheme also opens up possibilities for flexible manipulation of transport optics at arbitrary locations in the beam line. Many limitations imposed by the localized matching paradigm are lifted as a result. The recipes discussed in this report allow users to quickly and systematically implement such manipulations.

With infrastructure set up for distributed matching, it makes good sense to combine it with an interpolation scheme, in which pre-computed results are used to provide fast and proven matching solutions online. The concept of interpolated solution is independent of distributed matching, but their combined effect can better enhance and expedite the way transverse phase space is controlled than each scheme alone. In such a scheme all matching solutions are known and proven, with computation-intensive tasks done even before the machine is turned on. All online beam and transport manipulations amount to inputting measurements to this solution database and interpolating for a predictable and proven distributed solution, within a well-controlled minimal amount of time.

A deterministic matching algorithm is developed with the aim of realizing distributed matching. It relies on integrating the optimal trade-off curve between competing constraints, and is thus a departure from conventional approaches. Apart from fulfilling the goal of realizing distributed matching, it has unique features as a stand-alone matching tool in its own right, and enjoys many advantages over conventional algorithms. Its application is not limited to matching, but can be extended to optimization of any competing objectives with analytic models.

Another important application of the trade-off algorithm involves establishing continuously variable machine states tracing out all intermediate solutions for optimal trade-off between two competing options. Examples of this include: Beam orbit vs phase space; Halo vs phase space; Allocation between X & Y emittances; Head vs tail matching into FEL undulator (Figure 10), etc.

# VII. APPENDIXES

## APPENDIX A. GENERALIZED MISMATCH FACTOR

We set out to show that generalized mismatch factor $\Phi$ as defined in equation (3) for arbitrary dimensions is always greater or equal to one, with equality corresponding to perfect matching of beam covariance to design.

The design and measured beam covariances are denoted by $\Sigma_D$ and $\Sigma_M$ respectively. As $\Sigma_D$ and $\Sigma_M$ are connected through symplectic transformations[4], we have

$$Det(\Sigma_D) = Det(\Sigma_M) . \quad (A1)$$

Being positive-definite and symmetric with positive eigenvalues by definition, the covariance $\Sigma_D$ can be diagonalized by a real matrix $D$ that is a representation of $SO(2N)$, the special orthogonal group, with $2N$ the dimensionality of phase space of interest,

$$D \cdot \Sigma_D \cdot D^T = S = \begin{pmatrix} s_1 & 0 & 0 & 0 \\ 0 & s_2 & 0 & 0 \\ 0 & 0 & s_3 & 0 \\ 0 & 0 & 0 & s_4 \end{pmatrix}, \quad (A2)$$

$$D \in SO(2N) ,$$

where explicit form with $N=2$ is given as example, with $s_1, \cdots s_{2N}$ real positive eigenvalues of $\Sigma_D$. Note $SO(2N)$ group property dictates that $D^T = D^{-1}$. We can further apply a diagonal rescaling to arrive at identity matrix

$$\sqrt{S}^{-1} \cdot D \cdot \Sigma_D \cdot D^T \cdot \sqrt{S}^{-1} = I ,$$

$$\sqrt{S}^{-1} = \begin{pmatrix} 1/\sqrt{s_1} & 0 & 0 & 0 \\ 0 & 1/\sqrt{s_2} & 0 & 0 \\ 0 & 0 & 1/\sqrt{s_3} & 0 \\ 0 & 0 & 0 & 1/\sqrt{s_4} \end{pmatrix}, \quad (A3)$$

with $\sqrt{S}^{-1}$ a shorthand for the matrix shown above. The inverse of $\Sigma_D$ is likewise diagonalized by element in $SO(2N)$, which by group property must exist and is simply $D$ itself.

$$\sqrt{S} \cdot D \cdot \Sigma_D^{-1} \cdot D^T \cdot \sqrt{S} = K \cdot \Sigma_D^{-1} \cdot K^T = I, \quad (A4)$$
$$K = \sqrt{S} \cdot D.$$

Now we have

$$Tr(\Sigma_D^{-1} \cdot \Sigma_M)$$
$$= Tr(K \cdot \Sigma_D^{-1} \cdot K^T \cdot {K^T}^{-1} \cdot \Sigma_M \cdot K^{-1}) \quad (A5)$$
$$= Tr(\sqrt{S}^{-1} \cdot D \cdot \Sigma_M \cdot D^T \cdot \sqrt{S}^{-1})$$

where we used (A4). Note the quantity $D \cdot \Sigma_M \cdot D^T$ is again real symmetric and can be diagonalized by real matrix $E$ of $SO(2N)$,

$$E \cdot D \cdot \Sigma_M \cdot D^T \cdot E^T = T = \begin{pmatrix} t_1 & 0 & 0 & 0 \\ 0 & t_2 & 0 & 0 \\ 0 & 0 & t_3 & 0 \\ 0 & 0 & 0 & t_4 \end{pmatrix}, \quad (A6)$$

$$E \in SO(2N) ,$$

where $t_1, \cdots t_{2N}$ are real positive eigenvalues of $E \cdot D \cdot \Sigma_M \cdot D^T \cdot E^T$, and hence of $\Sigma_M$, by property of the group, and (A5) becomes

$$Tr(\Sigma_D^{-1} \cdot \Sigma_M)$$
$$= Tr(E \cdot \sqrt{S}^{-1} \cdot E^T \cdot T \cdot E \cdot \sqrt{S}^{-1} \cdot E^T). \quad (A7)$$

We make the final observation that the expression inside the parenthesis is real symmetric and can be diagonalized by yet another real matrix $O$ of $SO(2N)$,

$$Tr(\Sigma_D^{-1} \cdot \Sigma_M)$$
$$= Tr(O \cdot E \cdot \sqrt{S}^{-1} \cdot E^T \cdot T \cdot E \cdot \sqrt{S}^{-1} \cdot E^T \cdot O^T)$$
$$= Tr(P) \quad (A8)$$
$$= \sum_{k=1}^{2N} p_k , \quad O \in SO(2N) ,$$

$$P = \begin{pmatrix} p_1 & 0 & 0 & 0 \\ 0 & p_2 & 0 & 0 \\ 0 & 0 & p_3 & 0 \\ 0 & 0 & 0 & p_4 \end{pmatrix}.$$

---

[4] Constant momentum transport is assumed here. Otherwise all covariances should be momentum-normalized first.

where $p_1, \cdots p_{2N}$ are real positive eigenvalues of $E \cdot \sqrt{S}^{-1} \cdot E^T \cdot T \cdot E \cdot \sqrt{S}^{-1} \cdot E^T$, and naturally

$$\prod_{k=1}^{2N} p_k = Det\left(E \cdot \sqrt{S}^{-1} \cdot E^T \cdot T \cdot E \cdot \sqrt{S}^{-1} \cdot E^T\right) = 1, \quad (A9)$$

where we used (A1), (A2) and (A6). Rewriting (A8) and (A9),

$$Tr(\Sigma_D^{-1} \cdot \Sigma_M) = \sum_{k=1}^{2N} p_k,$$

$$\prod_{k=1}^{2N} p_k = 1, \quad (A10)$$

$$p_k > 0, \quad k = 1, \ldots, 2N.$$

The last line is inherent in the positive-definite property of the covariance, preserved by $SO(2N)$. (A10) implies an inequality condition on $Tr(\Sigma_D^{-1} \cdot \Sigma_M)$ subject to a constraint, in the form of Lagrange multiplier:

$$\begin{cases} \nabla_{p_k}\left(\sum_{k=1}^{2N} p_k\right) = \lambda \cdot \nabla_{p_k}\left(\prod_{k=1}^{2N} p_k\right) \\ \prod_{k=1}^{2N} p_k = 1 \\ p_k > 0, \quad k = 1, \ldots, 2N \end{cases} \quad (A11)$$

where $\nabla_{p_k}$ is the gradient operator in the space of $p_1, \cdots p_{2N}$, and $\lambda$ the Lagrange multiplier. The solution to (A11) gives the extremum of $Tr(\Sigma_D^{-1} \cdot \Sigma_M)$ of (A10) subject to the constraint. It is understood that the solution is to be confined to the "quadrant" of all positive $p_k$'s. (A11) is trivially solved to give

$$\lambda = 1; \quad p_k = 1, \quad k = 1, \ldots, 2N. \quad (A12)$$

This echoes the well-known inequality between algebraic and geometrical means of positive numbers,

$$\frac{1}{2N}\left(\sum_{k=1}^{2N} p_k\right) \geq \left(\prod_{k=1}^{2N} p_k\right)^{\frac{1}{2N}}, \quad (A13)$$

$$p_k > 0, \quad k = 1, \ldots, 2N$$

with equality corresponding to identical $p_k$'s. So the minimum of $Tr(\Sigma_D^{-1} \cdot \Sigma_M)$ happens when

$$Tr(\Sigma_D^{-1} \cdot \Sigma_M) = 2N,$$
$$P = I \quad (A14)$$
$$= O \cdot E \cdot \sqrt{S}^{-1} \cdot E^T \cdot T \cdot E \cdot \sqrt{S}^{-1} \cdot E^T \cdot O^T.$$

Now we can undo by steps the layers of transformation imposed on $\Sigma_M$. Using (A6) it is easy to get

$$O^T \cdot E^T \cdot I \cdot E \cdot O = \sqrt{S}^{-1} \cdot D \cdot \Sigma_M \cdot D^T \cdot \sqrt{S}^{-1}$$
$$\rightarrow I = \sqrt{S}^{-1} \cdot D \cdot \Sigma_M \cdot D^T \cdot \sqrt{S}^{-1} \quad (A15)$$
$$\rightarrow D \cdot \Sigma_M \cdot D^T = S.$$

Comparison of (A2) and (A15) shows this means

$$\Sigma_M = \Sigma_D, \quad (A16)$$

as a <u>necessary and sufficient</u> condition for $\Phi = Tr(\Sigma_D^{-1} \cdot \Sigma_M)/(2N) = 1$. Note this result comes mainly as a consequence of (A1) and the covariance properties of $\Sigma_D$ and $\Sigma_M$.

The transformation that $\Sigma_M$ undergoes in (A5) is akin to the transformation to normalized phases space defined by design beam in the 2D case. In more general situations $\Sigma_D$ and $\Sigma_M$ correspond to convex ellipsoids with equal volume by symplecticity but unmatched axial lengths and orientations. Action of $SO(2N)$ rigidly rotates these ellipsoids without deforming them. In the <u>normalized</u> space the only way for $\Sigma_D$ and $\Sigma_M$ to be matched is for both to be spheres of radius one, namely, with all axes having equal length. It is easy to see that in the 2D case ($N=1$) $Tr(\Sigma_D^{-1} \cdot \Sigma_M)/(2N)$ reduces to the familiar equation (1):

$$\Phi = (\beta_D \cdot \gamma_M - 2 \cdot \alpha_D \cdot \alpha_M + \gamma_D \cdot \beta_M)/2 \geq 1$$

# APPENDIX B. TRADE-OFF CURVE FOR COMPETING OBJECTIVES

Matching algorithm in beam optics has been a subject commanding lasting interest [2,15]. Below we present an alternative perspective to matching.

Consider two objective functions $F(k_m)$ and $H(k_m)$ of variables $k_m$ in a constrained optimization with $F$ the objective and $H=h$ the constraint. Optimum $f$ of $F$ subject to $H=h$ is determined by

$$\left. \begin{array}{c} \nabla F = \lambda \cdot \nabla H \\ H = h \end{array} \right\} \to F = f \ . \quad (B1)$$

The first relation above can be called the tangency condition, based in the picture of Lagrange multiplier. As $h$ varies, the optimal solution $f(h)$ varies as a function of $h$. The roles of $F$ and $H$ can be switched

$$\left. \begin{array}{c} \nabla H = \mu \cdot \nabla F \\ F = f \end{array} \right\} \to H = h \ , \quad (B2)$$

with $\mu = 1/\lambda$. Rather than being artificially demarcated as objective vs constraint, $F$ and $H$ can be viewed as competing objectives, and we can alternatively take as independent variable $\lambda$ or $\mu$, parametrizing a "trade-off" curve in the space of $k_m$ on which the tangency condition (B1) or (B2) is always satisfied. $F$ and $H$ take on values as functions of $\lambda/\mu$: $f(\lambda/\mu)$ and $h(\lambda/\mu)$, along this curve. Following equations (B1) and (B2) we have

$$\left. \frac{df}{dh} \right| = \lambda \ , \quad \left. \frac{dh}{df} \right| = \mu \ , \quad (B3)$$

where the vertical bar indicates that the derivatives are taken along the 1D locally optimal trade-off curve. The trade-off curve traces out a continuous path in the space spanned by $k_m$, on which each point represents the optimal local trade-off between $F$ and $H$ for a given $f$ or $h$, or equivalently, $\lambda$ or $\mu$. It terminates on both ends at extrema $F=f_0$ and $H=h_0$:

$$\begin{array}{c} \left. \nabla F \right|_{F=f_0} = 0 \ , \\ \left. \nabla H \right|_{H=h_0} = 0 \ , \end{array} \quad (B4)$$

corresponding to ($\lambda=0$; $\mu=\pm\infty$) and ($\mu=0$; $\lambda=\pm\infty$) respectively[5]. In the following we limit the discussion to cases where both $F=f_0$ and $H=h_0$ represent local minima, and the trade-off curve ends at ($\lambda=0$; $\mu=-\infty$) and ($\mu=0$; $\lambda=-\infty$) on both ends. The term "trade-off" is clear in the sense that since $\lambda$ and $\mu$ are both non-positive over the entire curve, any local gain in $F$ must be made at the expense of $H$ and vice versa. The trade-off between $F$ and $H$ is monotonic over the entire range.

With knowledge of $k_m$'s at any particular point on the trade-off curve, one can in principle trace the curve to either end and arrive at optima of $F$ and $H$ by integrating a differential form of the tangency condition. Equation (B3) alone however is not sufficient for this purpose for two reasons: a): there is no clear indication of when the integration should end, namely, when the true extremum has been reached, even by looking at the local behaviors of $f(\lambda)$ or $h(\lambda)$, and b): equation (B3) governing $f$–$h$ dependence is potentially singular. The latter point is subtle but related to the fact that $\lambda$ is always negative between the end points (B4), thus any non-monotonic bend or loop in the $f$–$h$ curve must do it through sharp spikes.

The correct way to integrate the trade-off curve can be worked out by bringing in $\lambda$ itself as the alternate variable,

$$\left. \frac{d\mathbf{k}}{d\lambda} \right| = \mathbf{M}^{-1} \cdot \mathbf{R} \ , \quad \mathbf{k} = \left( k_1^O(\lambda), k_2^O(\lambda), \ldots k_N^O(\lambda) \right),$$

$$\mathbf{M}_{ij} = \frac{\partial^2 (F(\mathbf{k}) - \lambda \cdot H(\mathbf{k}))}{\partial k_i \partial k_j}, \quad \mathbf{R}_i = \frac{\partial H(\mathbf{k})}{\partial k_i},$$

$$(B5)$$

where $k_m^O$'s are functions of $\lambda$ representing local optimal solutions at every point along the curve. Integrating (B5) alone is still insufficient, as it encounters singularities at

$$Det(\mathbf{M}) = 0 \ , \quad (B6)$$

upon which the evaluation of $\mathbf{k}$ must be continued by integrating over $f$ or $h$ instead:

---

[5] The sign depends on whether we are dealing with extrema of same or opposite flavor on both ends.

$$\left.\frac{d\boldsymbol{k}}{df}\right| = \frac{1}{\lambda} \cdot \frac{\boldsymbol{M}^{-1} \cdot \boldsymbol{R}}{\boldsymbol{R}^T \cdot \boldsymbol{M}^{-1} \cdot \boldsymbol{R}} = \frac{1}{\lambda} \cdot \frac{Adj(\boldsymbol{M}) \cdot \boldsymbol{R}}{\boldsymbol{R}^T \cdot Adj(\boldsymbol{M}) \cdot \boldsymbol{R}},$$

$$\left.\frac{d\boldsymbol{k}}{dh}\right| = \frac{\boldsymbol{M}^{-1} \cdot \boldsymbol{R}}{\boldsymbol{R}^T \cdot \boldsymbol{M}^{-1} \cdot \boldsymbol{R}} = \frac{Adj(\boldsymbol{M}) \cdot \boldsymbol{R}}{\boldsymbol{R}^T \cdot Adj(\boldsymbol{M}) \cdot \boldsymbol{R}}, \quad (B7)$$

$$Adj(\boldsymbol{M}) = Cof(\boldsymbol{M})^T = Det(\boldsymbol{M}) \cdot \boldsymbol{M}^{-1}$$

where $Adj(\boldsymbol{M})$ is the adjugate, or transpose of cofactor, of $\boldsymbol{M}$. Integration of (B7) can again encounter singularities when a saddle-point type condition is reached

$$\boldsymbol{R}^T \cdot Adj(\boldsymbol{M}) \cdot \boldsymbol{R} = 0, \quad (B8)$$

upon which the integration must revert back to (B5), and integration of the trade-off curve proceeds according to a recipe outlined as follows (assuming the goal is to obtain minimum for $F$, $\nabla F = 0$, from a known minimum for $H$, $\nabla H = 0$)[6].

B1. Start from the known optimum for $H$, corresponding to $\nabla H = 0$ and $\lambda = -\infty$,
B2. Integrate (B5) until (B6) occurs,
B3. Switch to one of (B7) until (B8) happens,
B4. Alternate between steps B2. & B3.,
B5. <u>Process terminates when $\lambda=0$</u>, where optimum $\nabla F = 0$ is reached on the other end of the curve.

The path traced out by the above integration, parametrized by $\boldsymbol{k} = \left(k_1^O(\lambda), k_2^O(\lambda), \ldots k_N^O(\lambda)\right)$ from $\lambda = -\infty$ to $\lambda = -0$, consists of all locally optimal trade-off solutions between $\nabla F = 0$ and $\nabla H = 0$. In the context of distributed matching, any point on this curve can be taken as an optimal partial matching solution towards the eventual full match. The above formulation is explored in-depth in the following, with particular attention paid to its effect on distributed matching.

### 1. Determinism

The point $\lambda=0$ marks the end of <u>monotonic</u> trade-off between $F$ and $H$. Beyond this point increased $H$ will lead to <u>increased</u>, thus less optimal, values for $F$. Step B5 above thus provides an unequivocal termination criterion, and a measure of determinism, to the process. The formulation presented here enjoys a level of determinism distinct from algorithms conventionally employed in accelerator matching in the following ways

- The process has a deterministic starting point dictated by the competing objective $H$. There is no need for <u>case by case</u> inspired guesses or random number searches, or artificial weighting between incongruent parameters.
- A rigid recipe is followed. There is no room for artificial parameter tweaking to guide the solution.
- The process continues and ends on unambiguous criteria:
  o If $\lambda=0$, stop,
  o If $\lambda\neq 0$, do not stop.

Such criteria are lacking in a conventional algorithm, often causing it to either be uncertain about when to stop, or stop short of more globally rewarding solutions. This will be made clear in examples below.

As this algorithm eliminates the need for case-by-case guesswork and tweaking, and as a result has much more smooth dependence on input variables, it possesses properties useful for automated or large-scale applications discussed in the main text.

### 2. Optimality and Global Properties

The recipe outlined in steps B1–B5 guarantees, under any condition, a solution with a well-defined meaning for its optimality: It is the best optimum (minimum in the matching context) attainable for $F$ while staying within the regime of monotonic trade-off with $H$. The monotonic trade-off, enforced by the condition $\lambda\leq 0$, ensures that for any given $F=f$, the corresponding point on the curve has the lowest $H=h$, and vice versa. More complicated multi-valued cases require a procedure akin to Pareto front isolation in optimization and will be discussed later. An example of application of Pareto front isolation to accelerator modeling is found in [3]. Figure B1(a) shows how the final solution for $\nabla F = 0$ depends on the choice of competing objective, $H$ or $H'$, by ensuring that whichever one is chosen remains locally optimized at every point on the respective trade-off curve. The final solution for $\nabla F = 0$ thus bears the imprint of the original constraint in the desirable way.

---

[6] Also see conjugate formulation below (Appendix Appendix B4) for avoiding infinity in the first step.

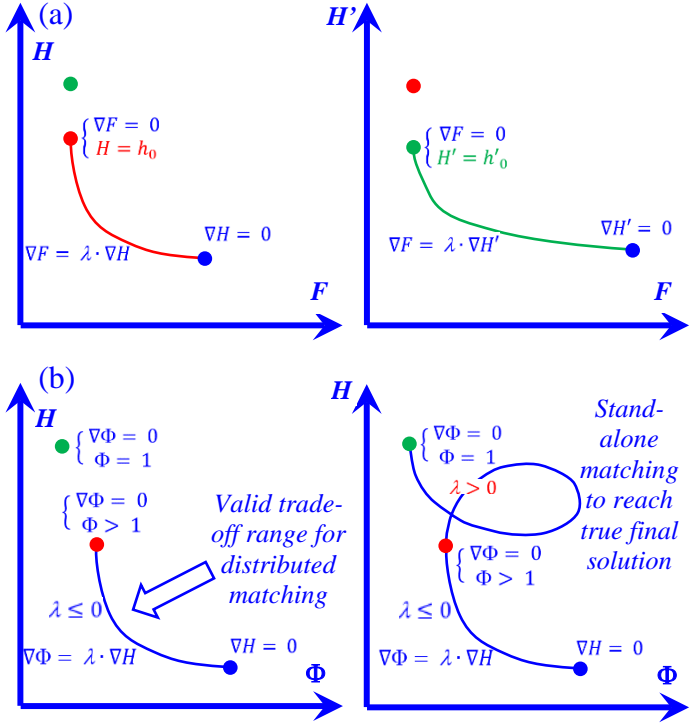

Figure B1. Properties of the trade-off curve derived by integrating (B5) and (B7). (a): Optimal solution $\nabla F = 0$ at the end of integration depends on choice of competing objective $H$. The two plots show integration paths starting from different $H$ and $H'$. In each case the end point $\nabla F = 0$ lands on the solution corresponding the smallest competing objective of interest. Red (Green) dots in both plots represent the same $\nabla F = 0$ solution when $H$ ($H'$) is used as the constraint. (b): Condition to stop at $\lambda=0$ always produces the entire stretch of solutions with monotonic trade-off between $\nabla \Phi = 0$ and $\nabla H = 0$. Monotonic trade-off is not valid beyond this point (red dot), although for the purpose of 100% matching, or extreme trade-off as discussed in Section V, one can integrate past the red dot to reach further points of $\nabla \Phi = 0$ (green dot).

In the context of matching, we use the mismatch factor $\Phi$ as defined in (3) or (6) for one of the objectives, $F$, and the quadratic sum of deviations from design quadrupole strengths, $\Delta K$ as in (8), for the competing objective $H$. In over-constrained (e.g., using 3 quadrupoles) or even some critically constrained (4 quadrupoles) matching cases, there can be no exact matching solutions with $\Phi=1$ [2]. In such cases the recipe gives an unambiguous path to the best solution as $\lambda$ reaches 0, a result that conventional algorithms cannot deliver.

In rare occasions, especially in critically constrained cases, the trade-off curve can encounter $\lambda=0$ without $\Phi=1$ while an exact solution $\Phi=1$ does exist. In other words, the process terminates at a false local minimum despite existence of true minima. This point is still the best solution within the regime of monotonic trade-off, beyond which the integration (B5) and (B7) enter the realm of simultaneously degrading performance ($\lambda>0$) for both $\Phi$ and $H$. For distributed matching where the goal is to find the best trade-off between $\Phi$ and $H$ without necessarily reaching $\Phi=1$, the relevant section of trade-off curve is already obtained and as valid a final result as in any other case. On the other hand, if the algorithm is used as a stand-alone matching tool and the goal is to find the true solution $\Phi=1$, one would have to integrate beyond this point until the true $\Phi=1$ solution is reached, often at a much stiffer penalty in $H$ (e.g., $\Delta K$). This concept is depicted in Figure B1(b).

More extreme trade-off can be achieved if one continues on the integration path, which will lead to further optimal solutions $\Phi=1$, but at progressively increasing cost to $H$. Figure B2 shows a realistic 4-quadrupole matching example in a 30° FODO lattice, where the competing objective is taken to be the quadratic sum of quadrupole strengths $H=K^2$. The regime of optimal trade-off between $\Phi$ and $K^2$ terminates at the first $\lambda=0$ and $\Phi>1$ (magenta point), beginning to approach the point of diminished return (Appendix B5). If the goal is however to achieve full matching at all cost, the integration can continue into the region of $\lambda>0$ until, after significant increase in $K^2$, a true solution with $\lambda=0$ and $\Phi=1$ is reached (green dot). This solution is nevertheless globally the best full matching solution subject to the competing objective $H(=K^2)$. In other words, it is the full matching solution of $\Phi=1$ with the smallest $K^2$. Even more severe penalty in $K^2$ is incurred if one continues on with the integration to reach the next full matching solution, and so on. In practice this continuation into ever more inferior solutions can serve a practical purpose by deliberately mismatching against a competing objective, as explained in the last example of Section V.

The power of determinism as implied in the recipe may be better appreciated with a realistic example of matching in a 6-quadrupole 30° per cell FODO lattice as shown in Figure B3. Once the process starts, one only needs to blindly adhere to the recipe, never stopping until $\lambda=0$, even if $\Phi$ gets to within $10^{-4}$ of unity at intermediate points[7], or if the process seems to lead farther afield of the goal in the interim. In the end one is rewarded with a true solution <u>globally</u> superior to more premature solutions had the process been terminated early. This final solution is superior not only in $\Phi$, but also in the competing objective $\Delta K$.

### 3. Pareto front isolation

Figure B3 indicates that there are cases where the trade-off curve is multiple valued for a given intermediate objective. Although at every point of the integration monotonic trade-off is locally satisfied ($\lambda \leq 0$), globally there can be counter examples to monotonic trade-off across different "branches" of the curve. This is undesirable for the purpose of a deterministic process, such as distributed matching, in which a <u>unique</u> optimal solution must be identified for any given objective in either $F$ or $H$. In order to restore this determinism, we resort to the concept of Pareto front [3] in multi-objective optimization, where a subset of the solution ensemble is isolated in which none of the solution is "dominated", namely, inferior in terms of the <u>entire</u> set of competing objectives, by any other solution in the set. Thanks to the fact that $\lambda \leq 0$ everywhere, such an isolation can be done in an unambiguous way. The trade-off curve Figure B3(i), reproduced in Figure B4(a), can thus be separated at the branch intersection point (green dot) into the Pareto front (thick green line) and the inferior solution set (thin blue line) shown in Figure B4(b). After this process only globally superior points remain. The Pareto front isolation restores <u>global</u> <u>monotonic</u> trade-off in the solution curve, in the sense that for any given $F$, this curve again gives the unique global minimal $H$, and vice versa. Figure B4(c) shows evolution of individual $k_m$'s (quadrupole strength) by trade-off integration (B5) and (B7), color correlated with B4(a). Figure B4(d) shows the consequence of Pareto front isolation on individual $k_m$'s. A discontinuity is introduced, short-circuiting each $k_m$ path, with the welcome effect of cleaning up an otherwise convoluted trajectory for the $k_m$'s.

### 4. Conjugate formula, initial slope, and derivative relations

Formulas conjugate to (B5) and (B7) can be introduced ($\mu = 1/\lambda$) to circumvent difficulty from $\lambda=-\infty$ at the starting point (Step B1) of the integration recipe:

$$\left.\frac{d\bm{k}}{d\mu}\right| = \bm{N}^{-1} \cdot \bm{S} , \qquad \bm{k} = \left(k_1^O(\mu), k_2^O(\mu), \ldots k_N^O(\mu)\right),$$

$$\bm{N}_{ij} = \frac{\partial^2 \left(H(\bm{k}) - \mu \cdot F(\bm{k})\right)}{\partial k_i \partial k_j}, \qquad \bm{S}_i = \frac{\partial F(\bm{k})}{\partial k_i} .$$

(B9)

Thus step B1 in the integration recipe B1–B5 above can be preceded by an extra step:

B0. Start from the known optimum for $H$, corresponding to $\nabla H = 0$ and $\mu=0$, and integrate (B9) towards <u>negative</u> $\mu$. As $\mu$ evolves into a finite negative value by integration, set $\lambda = 1/\mu$ and proceed to step B1 above.

To integrate (B9) from $\nabla H = 0$ one needs to know the local derivative at $\mu=0$. To do this the explicit form of $H$ is needed. Taking the form of $H$ defined in equation (7) or (8), this initial derivative can be calculated from

$$\bm{N}_{ij}\big|_{\mu=0} = \frac{\partial^2 H(\bm{k})}{\partial k_i \partial k_j} = 2\delta_{ij}$$

$$\rightarrow \left.\frac{d\bm{k}_i}{d\mu}\right|_{\substack{\mu=0 \\ k_m=0}} = \frac{1}{2}\left.\frac{\partial F(\bm{k})}{\partial k_i}\right|_{k_m=0}$$

(B10)

Note that $h, f, \lambda$ and $\mu$ are functions of each other along the trade-off curve as dictated by Table B1.

### 5. Points of diminished return and inflection in trade-off curve

It is obvious from the example in Figure B5 that the rate of trade-off between competing objectives is not a constant across the curve, on which one can discern a "point of diminished return" beyond which the gain in one objective ($\Phi$) hardly justifies the requisite cost in the other ($\Delta K$). Given the global behavior of the curve computed by the algorithm, criteria and recipe can be developed to define such points, in distributed matching for example, to guide selection of optimal intermediate matching solutions for the "best bang for the buck". Such freedom is possible only in distributed matching, as in local schemes full matching on the spot is the only option even if it translates into heavy expense on quadrupole strengths to cover the last few percent of mismatch, of which the algorithm may not even be aware. The algorithm presented here provides a continuous set of options and a picture of global trade-off from which the user can choose the optimal one based on special criteria at hand.

---

[7] In numerical testing on the same problem a conventional optimization-based matching algorithm indeed returned such inferior solutions as the final answer. The numerical example of Figure B3 shows that such solution is the <u>one to be avoided</u>, requiring more than twice the quadrupole strength deviation!

We will not delve into criteria and recipe for identifying such points in this report, other than noting a special case where a well defined analytical condition can serve as a guide. From Table B1 it is easy to see

$$\left.\frac{d^2 f}{dh^2}\right| = \left.\frac{d\lambda}{dh}\right| = \frac{1}{\boldsymbol{R}^T \cdot \boldsymbol{M}^{-1} \cdot \boldsymbol{R}} \ . \tag{B11}$$

Therefore one can identify points of inflection on the trade-off curve by the condition making (B10) vanish, which is simply (B6), plus the condition for local minimum for $\lambda$,

$$Det(\boldsymbol{M}) = 0 \ ,$$
$$\left.\frac{d\lambda}{dh}\right| = 0 \ , \quad \left.\frac{d^2\lambda}{dh^2}\right| > 0 \ . \tag{B12}$$

Such condition marks points separating accelerated and decelerated trade-offs between competing objectives. In some cases this can be a candidate for the point of diminished return. Figure B5 shows stages of applying the algorithm to the matching in a 6-qudrupole system, with initial $\Phi$ exceeding 7200 when competing objective K, as defined in (7), is 0. Pros and cons of several intermediate solutions are obvious from Figure B5, including points at or following the inflection point (B11) shown in Figure B5(d). Despite the wide range of options, a localized scheme can only use the last solution in Figure B5 even though it is not the best use of quadrupoles.

It is also worth noting that tolerance on implementing quadrupole solution goes in the opposite direction, as is intuitively obvious. The required accuracy for implementing matching solutions is more stringent before the point of diminished return than it is after.

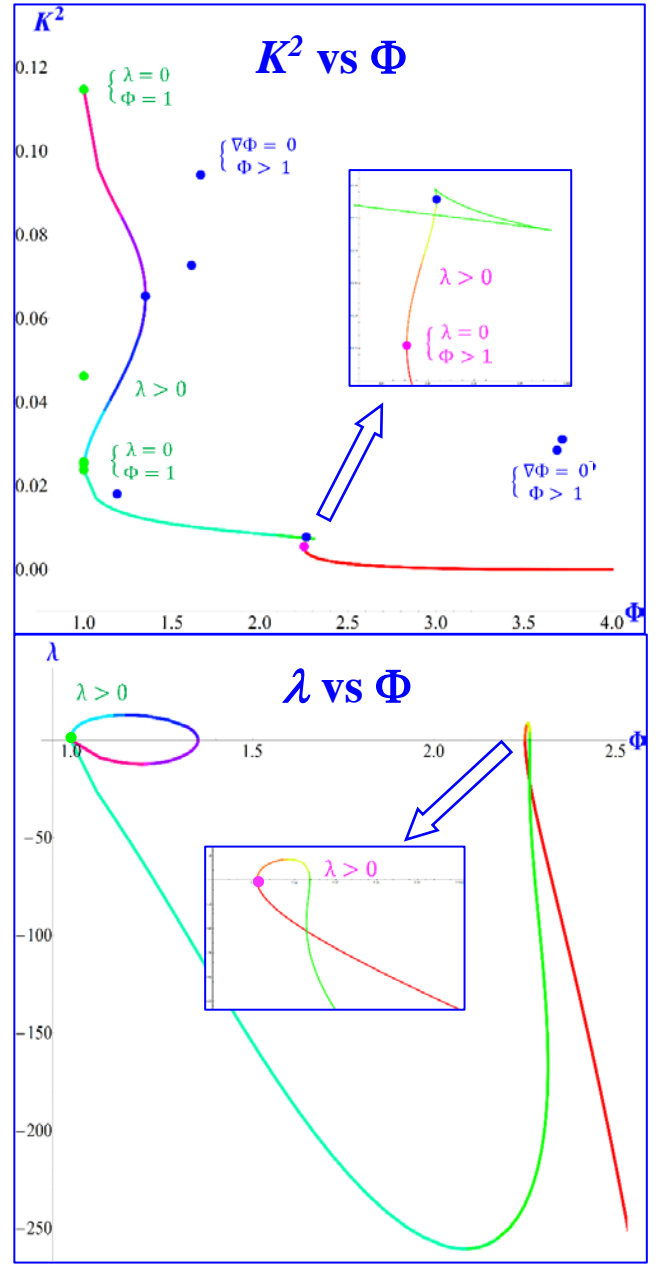

Figure B2. Realistic example of extreme trade-off on 4 quadrupoles in a 30° lattice. Both plots show different views of the same integration path from lower right (red) to upper left (purple). Corresponding segments are color correlated. Insets contain details. Initial monotonic trade-off between quadrupole strength ($K^2$) and $\Phi$ terminates at magenta dot when $\lambda$ first reaches 0 ($K^2$=0.01, $\Phi$=2.25). This (red) segment alone gives the entire solution set needed for distributed matching. If the algorithm is instead intended as a matching engine, integration can continue into territory of $\lambda$>0 (orange and yellow) and reach, after looping, the true solution at $\Phi$=1 (green dot, $K^2$=0.025) with penalty in $K^2$. More extreme trade-off is obtained by continuing on integration path through more looping, reaching $\Phi$=1, $K^2$=0.115, a severe penalty that may be useful for other purposes (See Section V).

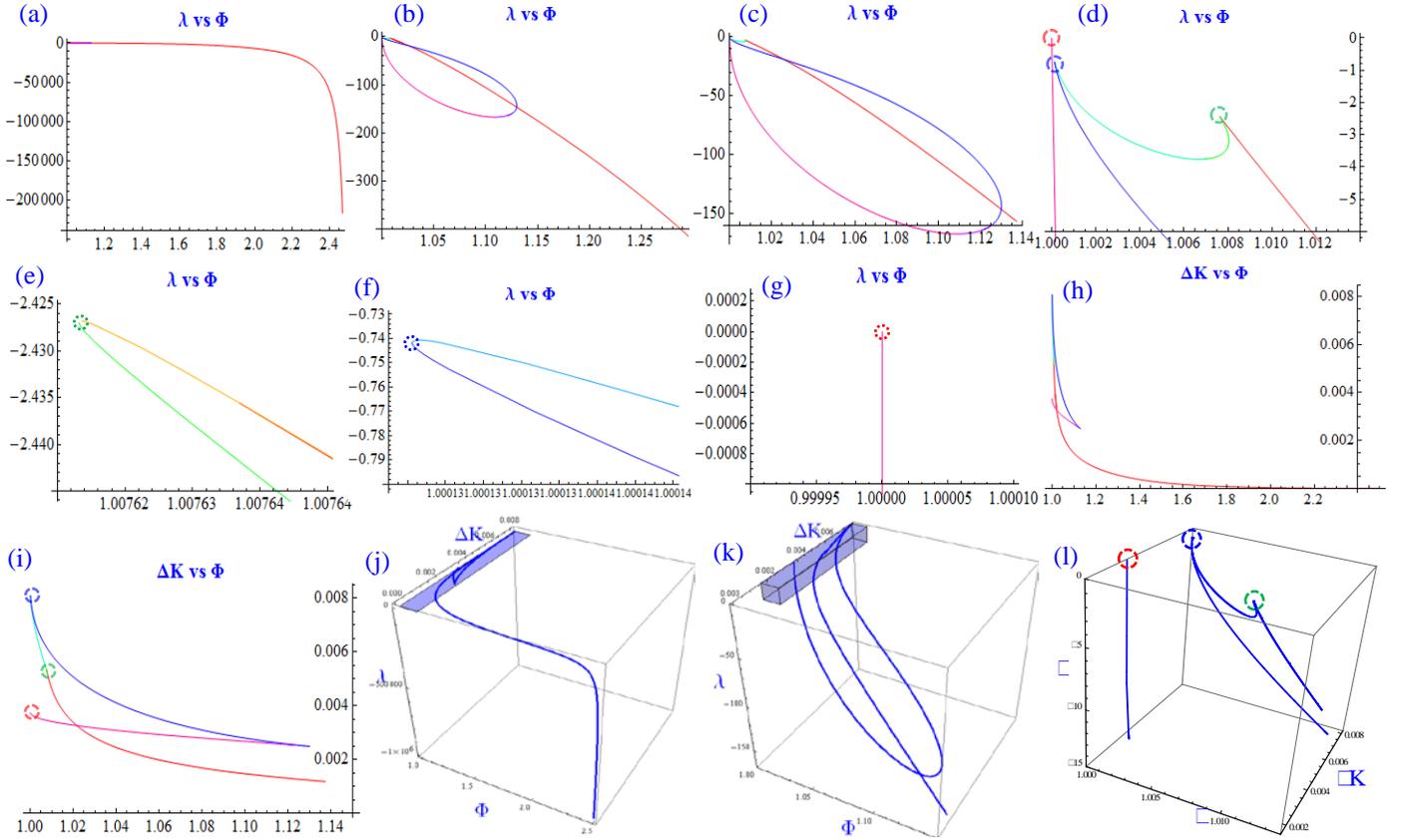

Figure B3. (a)–(g): Same integration path from $\lambda \to -\infty$ to $\lambda=0$, zoomed into different regions. The path approached $\Phi=1$ twice, indicated by green and blue circles, at $\Phi=1.0076$ & $1.00013$ (e-f), before looping all the way back to $\Phi=1.14$ (c) and returning to $\Phi=1$ (red circle). Conventional algorithm can stop short of the true optimum and return $\Phi=1.00013$ as final answer. (h)–(i): Same trade-off curve zoomed into different regions. Superiority of the true $\Phi=1$ solution over $\Phi=1.00013$ is obvious in its much smaller quadrupole strengths. Corresponding segments in (a)–(i) are color correlated. (j)–(l): Same integration path from $\lambda \to -\infty$ to $\lambda=0$, zoomed into different regions (shaded in each preceding graph) in the space spanned by $\lambda$, $\Phi$, and $\Delta K$. The green, blue and red circles mark the same false and true $\Phi=1$ solutions.

### 6. Singularity-free formulation

Formulation (B5) and (B7) are useful in illustrating the properties of the trade-off curve, although the need to switch integration formula upon singularity makes the recipe cumbersome. A completely singularity-free formulation can be established by focusing on the instantaneous direction cosines of the trade-off curve in the space of $k_m$'s. This leads to

$$\frac{d\mathbf{k}}{dk} = \pm \hat{\mathbf{P}}, \qquad \mathbf{P} = Adj(\mathbf{M}) \cdot \mathbf{R}, \qquad (B13)$$

where we shorthanded $|d\mathbf{k}|$ as $dk$, bold faced letters $\mathbf{k}$ and $\mathbf{P}$ denote vectors, and caret denotes unit vector. Sign ambiguity is necessary and should be resolved by the flavor (maximum or minimum) of the end points.

Since $\mathbf{M}$ depends on $\lambda$, the integration must evaluate $\lambda$ as well at each step:

$$\frac{d\lambda}{dk} = \pm \frac{Det(\mathbf{M})}{|\mathbf{P}|},$$

$$\frac{df}{dk} = \pm \frac{\lambda \cdot (\mathbf{R}^T \cdot Adj(\mathbf{M}) \cdot \mathbf{R})}{|\mathbf{P}|} = \pm \lambda \cdot \mathbf{R}^T \cdot \hat{\mathbf{P}}, \qquad (B14)$$

$$\frac{dh}{dk} = \pm \frac{(\mathbf{R}^T \cdot Adj(\mathbf{M}) \cdot \mathbf{R})}{|\mathbf{P}|} = \pm \mathbf{R}^T \cdot \hat{\mathbf{P}}.$$

The signs chosen in (B14) should be the same as that in (B13). A zero on the right-hand side of any of the above equations marks a "turn-around" in the previous formulation without introducing singularity. The process stops when $\lambda=0$. Demand to avoid initial infinite $\lambda$ leads to the conjugate expression with $\mu = 1/\lambda$ as before:

$$\frac{d\mathbf{k}}{dk} = \pm\hat{\mathbf{Q}}, \quad \mathbf{Q} = Adj(\mathbf{N}) \cdot \mathbf{S},$$

$$\frac{d\mu}{dk} = \pm\frac{Det(\mathbf{N})}{|\mathbf{Q}|},$$

$$\frac{df}{dk} = \pm\frac{(\mathbf{S}^T \cdot Adj(\mathbf{N}) \cdot \mathbf{S})}{|\mathbf{Q}|} = \pm\mathbf{S}^T \cdot \hat{\mathbf{Q}}, \quad (B15)$$

$$\frac{dh}{dk} = \pm\frac{\mu \cdot (\mathbf{S}^T \cdot Adj(\mathbf{N}) \cdot \mathbf{S})}{|\mathbf{Q}|} = \pm\mu \cdot \mathbf{S}^T \cdot \hat{\mathbf{Q}}.$$

In this formulation the initial slope (B10) becomes

$$\left.\frac{d\mathbf{k}}{dk}\right|_{\mu=0,\, k_m=0} = \pm\hat{\mathbf{T}}, \quad \mathbf{T} = \nabla_{\mathbf{k}} F(\mathbf{k})\big|_{k_m=0}. \quad (B16)$$

This is intuitively obvious, as the initial direction for the motion of $\mathbf{k}$ should coincide with the gradient of $F$. The best-conditioned strategy is to start with (B15) and (B16) at $\mu=0$, integrate until $\mu = \lambda = -1$, then switch to (B13)–(B14) until the end, $\lambda=0$. This formulation can be rigorously shown to be singularity-free, obviating need to monitor or circumvent singularities like (B6) or (B8). This is graphically shown in Figure B6.

*Systematic recipe for obtaining family of solutions in under-constrained cases*

A special case worth discussing happens when the system is under-constrained, such as matching using 5 quadrupoles. When $\lambda=0$, the matrix $\mathbf{M}$ of (B5) becomes rank-deficient because $\mathbf{F}$ (i.e., $\Phi$) is a 5×5 matrix with only 4 degrees of freedom afforded by the 4D mismatch parameter, and thus $d\lambda/dk = 0$ by (B14) and $\lambda$ will stay 0 all the way. In the meantime the integration of $\mathbf{k}$ of (B13) will trace out a 1-dimensional space in $\mathbf{k}$ which forms the solution space for the matching problem. This provides a rigorous and systematic recipe for obtaining the complete family of matching solutions in an under-constrained problem, and can be useful in applications such as establishing phase-trombones where a system of 6 or more quadrupoles are used to hold the Twiss parameters fixed while varying betatron phases. In the current example the range of the phase trombone phase is spanned by the 1D or 2D family of solutions. Such results may be not easily obtained via a traditional matching algorithm.

## 7. Comprehensive numerical testing

The algorithm has been subjected to comprehensive numerical testing to evaluate its robustness. Given its deterministic nature, this algorithm should encounter no "difficult" matching problem in the sense of not being able to find the right initial solution neighborhood, or to stop when the best solution has been achieved, or to know if a given solution is optimal with respect to constraints. The path to the optimal solution is practically guaranteed, with the only challenge coming from numerical precision. Numerical testing does indicate the number of significant digits can be critical to the robustness of the algorithm in some cases, which should not pose an insurmountable obstacle with modern algorithmic platforms such as Mathematica or Matlab, to name a few. It should be emphasized that while number of significant digits is important for the integration process to negotiate sensitive corners of the trade-off curve, it does not mean the same level of precision is required of the implementation of the final solution in the machine, which is much more relaxed.

Systematic tests were carried out on the basic configuration shown in Figure B7, where a FODO lattice with $N_Q$ quads and phase advance $\psi$ per cell is used to rematch incoming beam covariance of mismatch amplitude $\Lambda$ and angle $\Theta$ in both planes (See Figure 6(a)). As explained in Figure 2, this addresses both beam matching and transport error correction. The cases studied include:

- $N_Q$: 1, 2, 3, 4, 5, 6, 7
- $\psi$: 30° – 150° at intervals of 30°
- $\Lambda_{X/Y}$: 1.2 – 8.0 at intervals of 0.2, 0.4 or 1.0
- $\Theta_{X/Y}$: 0° – 180° at intervals of 30° or 45°

The following observations are made:

*Over-constrained cases*

The over-constrained cases ($N_Q<4$), even when $N_Q=1$, are not as trivial as they appear, as when $\Phi\neq1$, the optimal combined X and Y match is not always obvious. Here the current algorithm displays a clear advantage over conventional methods, giving unambiguous condition on the optimum for $\Phi \neq 1$ and thus on when to terminate the process. Figure B8 shows the collection of outcome from a comprehensive scan over mismatch $\Lambda_{X/Y}$ and $\Theta_{X/Y}$ in a one, two three or four-quadrupole system in a 120° lattice, as well as a one or four-quadrupole system in a 30° lattice. These are already a prototype of the interpolation table proposed in Section IV3.

The difficulty to match based on a 30° lattice compared to 120° is visible, due to not cleanly decoupled $\beta_{X/Y}$. It is also worth noting that when $\Theta_{X/Y}$ are not in good orientation, a low initial $\Phi$ can be more difficult to match than a higher initial $\Phi$. Finally there are cases where an initial mismatch in only one plane can be more difficult than initial mismatch in both planes of similar $\Phi$.

In the case of 4 quadrupoles a real-valued solution does not always exist for a thin lens system [2], in which case the current algorithm can again give unequivocal direction on when to stop. In addition, the monotonic trade-off can stop short of a $\Phi=1$ solution. While this is sufficient for the purpose of distributed matching, if the true $\Phi=1$ solution is intended as the final answer, integration should be extended further to achieve this (Appendix B2 and Figure B2). This is not reflected in Figure B8.

*Correcting transport error within matching section*

A more stringent test of robustness is done where a transport error is introduced inside the matching section itself. This test is interesting in two ways: a). usually the mismatch factor thus introduced can be very large (several 1000), a good test of robustness, b). it is a good test of special optical module embedded within matching sections, as required for realistic implementation discussed in Section IV. To this end we inserted an additional "error" transfer matrix into each one of the 5 inter-quadrupole slots of a 6-quadrupole matching system and carried out the same trade-off integration as before, except that the overall transport $M(k)$ in (6) now contains the error transport as well. In the presence of embedded transport error the baseline quadrupole strength does not have much significance any more, thus we tested both scenarios of competing objective $H$: (7) and (8), namely, minimizing either absolute (K) or incremental ($\Delta$K) quadrupole strength. The resulting trade-off curve can take on very intricate patterns due to high non-linearity, especially at low phase advance. Figure B9 shows two examples of a 30° lattice with embedded error, one minimizing K and the other minimizing $\Delta$K. Despite the convoluted paths taken by the trade-off and the quadrupoles, after Pareto front isolation (Appendix B3) these paths will look much cleaner. For example in Figure B9(a) $k3$ remains almost constant near 0 after Pareto front isolation.

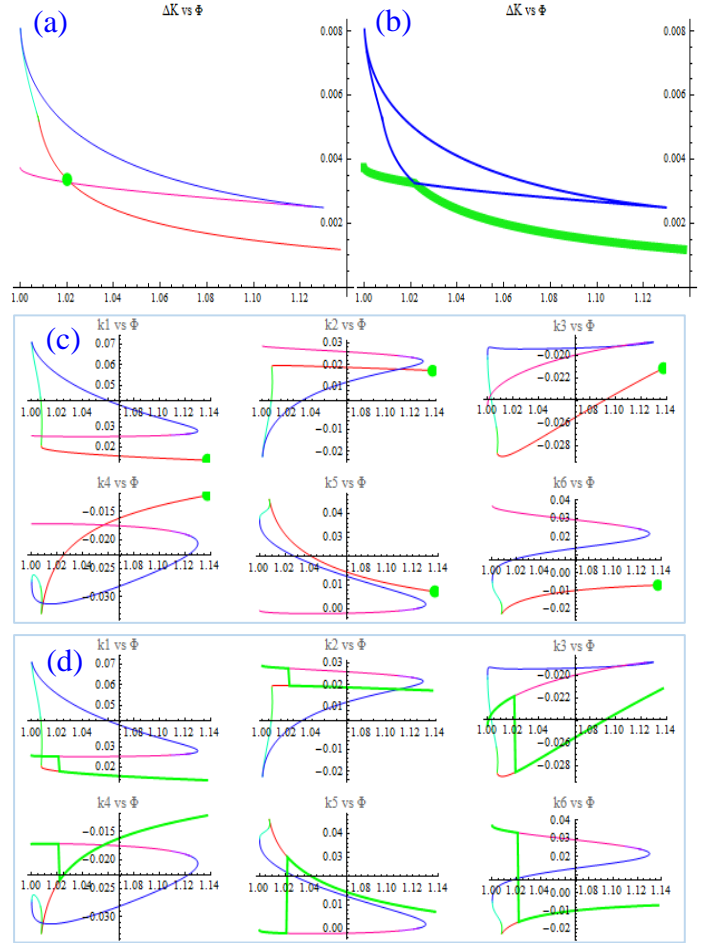

Figure B4. Extracting Pareto front from the solution curve for globally optimal trade-off. (a): Solution path in the $\Delta K$–$\Phi$ plane corresponding to Figure B3(i). Pareto front is extracted by joining the red and the magenta sections at the intersection (green dot), and discard everything above and to the right, as shown in (b). The fact that $\lambda$ is negative everywhere makes this extraction process unambiguous for both $\Phi$ and $\Delta K$. (c): Partial paths taken by the 6 quadrupole km's as functions of $\Phi$, color correlated with (a). (d): Effect of Pareto front extraction on quadrupole paths, each of which is short-circuited to follow the thick green path.

The claim made in Appendix B2 and Figure B1, about the solution biased toward minimizing whatever happens to be chosen as the competing objective $H$, is borne out in the same numerical tests. In Figure B10 are shown cases of matching with embedded errors. Depending on the choice of competing objective, either K or $\Delta$K, the final $\Phi$=1 solution is biased in favor of this choice. This is an expected but still welcome feature of the algorithm as it provides the user with an effective handle to steer the direction of the solution, including bias toward unconventional constraints, such as large mismatch in a different longitudinal slice of the beam as discussed in the last example of Section V2. The current algorithm has considerable advantage in this regard, especially in under-constrained systems ($N_Q$>4), where the solution space is multi-dimensional, and the use of an algorithm analytically ensuring this bias as the solution navigates this multi-dimensional space is of critical importance. For arguments given in Section IIIB and examples of Appendix B2, it is less obvious a conventional algorithm can always achieve this level of performance.

As mentioned earlier, there is no case-by-case guesswork or parameter tweaking required of all the above tests. Once adequate numerical precision is allocated, the computation proceeds to the end blindly, free of human intervention.

Determinism of the current algorithm hinges on one critical input, namely, knowledge of the (trivial) optimum $\nabla H = 0$ anchoring one end of the tradeoff curve, from which the integration recipe can be launched (Step B0 or B1 of the integration recipe). The a priori knowledge of this point, or actually of any point on the trade-off curve, is thus key to the algorithm's being deterministic. There will however be problems in a more complex situation where a point solution cannot be trivially known in advance. In such cases determinism will then not be a given.

A procedure may nonetheless overcome this obstacle and restore determinism. Assume the new goal is to solve the trade-off between two functions $F$ and $G$, and neither optimum, $\nabla F = 0$ or $\nabla G = 0$, is known a priori. This would not stop us if we notice, for example, $\nabla F = 0$ is a condition only dependent on $F$, so this point is a common terminus for trade-off curves between $F$ and all other functions. Nothing prevents us from artificially taking a trivial function such as $H = \Delta K$ of (7), integrating the trade-off first from $\nabla H = 0$ to $\nabla F = 0$, then onwards to $\nabla G = 0$ [8]. This is conceptually depicted in Figure B11. This possibility has the potential of extending the current algorithm to a much wider range of optimization problems for the control of other accelerator parameters and processes. One example of its usage is given in Section V2 of controlling differential mismatch between head and tail slices in a SASE FEL driver beam.

### 8. Restoring determinism without a priori knowledge of the trade-off curve

| $dA/dB|$ | $A=f$ | $A=h$ | $A=\lambda$ | $A=\mu$ |
|---|---|---|---|---|
| $B=f$ | 1 | $1/\lambda = \mu$ | $\left(\lambda \cdot \boldsymbol{R}^T \cdot \boldsymbol{M}^{-1} \cdot \boldsymbol{R}\right)^{-1}$ | $\left(\boldsymbol{S}^T \cdot \boldsymbol{N}^{-1} \cdot \boldsymbol{S}\right)^{-1}$ |
| $B=h$ | $1/\mu = \lambda$ | 1 | $\left(\boldsymbol{R}^T \cdot \boldsymbol{M}^{-1} \cdot \boldsymbol{R}\right)^{-1}$ | $\left(\mu \cdot \boldsymbol{S}^T \cdot \boldsymbol{N}^{-1} \cdot \boldsymbol{S}\right)^{-1}$ |
| $B=\lambda$ | $\lambda \cdot \boldsymbol{R}^T \cdot \boldsymbol{M}^{-1} \cdot \boldsymbol{R}$ | $\boldsymbol{R}^T \cdot \boldsymbol{M}^{-1} \cdot \boldsymbol{R}$ | 1 | $-\mu^2$ |
| $B=\mu$ | $\boldsymbol{S}^T \cdot \boldsymbol{N}^{-1} \cdot \boldsymbol{S}$ | $\mu \cdot \boldsymbol{S}^T \cdot \boldsymbol{N}^{-1} \cdot \boldsymbol{S}$ | $-\lambda^2$ | 1 |

Table B1. Derivative relations between $h$, $f$, $\lambda$ and $\mu$ along the 1D trade-off curve.

---

[8] It should be kept in mind that the trade-off between $F$ and $G$ thus arrived at will be a function of the choice of $H$. This is not necessarily a shortcoming, as one might be able to manipulate the property of the trade-off curve in his favor through judicious choice of $H$.

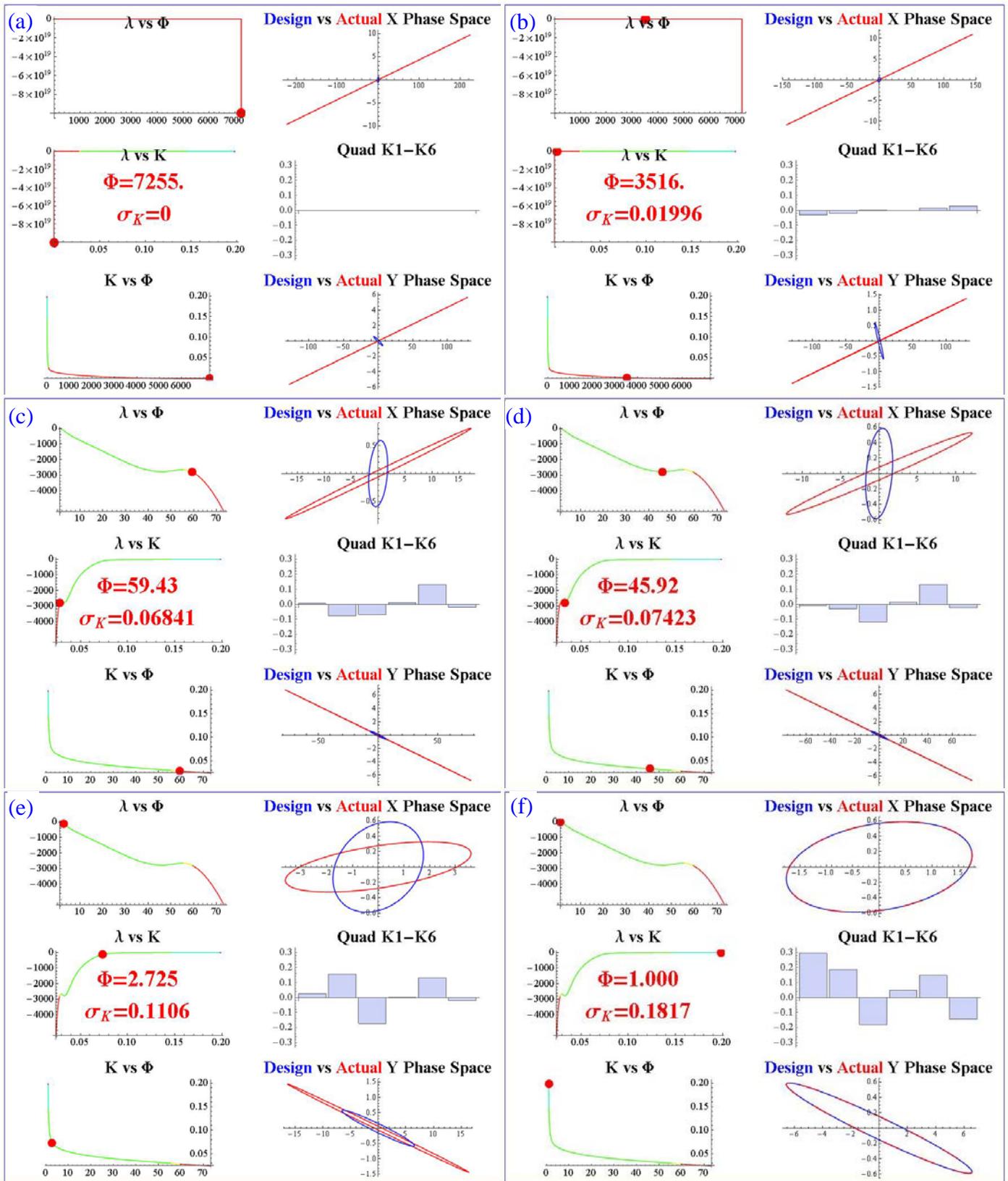

Figure B5. Six stages of 6-quadrupole matching in a 120° lattice. Competing objectives are mismatch Φ and quadratic sum K (m$^{-1}$) of inverse focal length. Each graph represents a point (red dot) in the solution path (red to green). Ellipses on the right represent target (blue) and intermediate (red) beams. Bar chart shows strength of all 6 quadrupoles. (a): Initial mismatch factor Φ=7255. Design ellipses are barely visible. (b-c): Small amount of correction reduced Φ to 59. (d): Inflection point ($Det(M)$=0, local minimum in $\lambda$) approaching <u>point of diminished return</u>. (e): Further up the curve past diminished return. (f): 100% matching. It took an increase of 0.11 m$^{-1}$ in RMS quadrupole strength ($\sigma_K$) to bring Φ down by 7250, but another 0.07 m$^{-1}$ only reduced Φ by 1.7!

### 9. Artificially guiding trade-off curve to global optimum

When used purely as a solution engine, the efficiency with which the algorithm homes in on the global optimum, $\nabla F = 0$, depends on two factors: a). proximity of target $\nabla F = 0$ to the starting point, and b). depth and steepness of $\nabla F = 0$ relative to lesser local optima, affecting how the trade-off curve is biased toward it. It may be possible to improve the efficiency and robustness of the algorithm by artificially enhancing the slope of the objective function $F$, for example, with an exponential amplifier,

$$F \to Exp(F), \qquad (B17)$$

effectively distorting the topography of the solution space and more actively steering the trade-off curve toward the global optimum. Such practice may enhance the resolution required to differentiate between true ($\Phi=1$) and false ($\Phi\approx1$) optima in the example given in Appendix B2 as well.

### 10. Solution by reversed beam path

With the picture of the current matching algorithm being one of following the path of locally optimal tradeoff from one quadrupole state (unmatched) to another (matched), a natural extension of this picture is to trace out the reversed beam path, with the beam line inverted in the sense of (16), where the initial quadrupole state brings the reversed target beam covariance $\Sigma_D^{Rev.}$ through the inverted beam line to a new reversed beam covariance $\Sigma_T^{Rev.}$ at the entrance, different from the reverse of the original initial beam covariance, $\Sigma_{IN}^{Rev.}$. Now the algorithm will need to follow the locally optimal path from $\Sigma_T^{Rev.}$ to $\Sigma_{IN}^{Rev.}$ for the inverted beamline. This will produce an equally legitimate alternative solution to the original problem.

### 11. Solution in under-constrained cases – A realistic phase trombone

As discussed in Section 6 earlier. This algorithm can be employed to perform systematic exploration of the solution space in an under-determined system. This is useful when it is desirable to scan additional parameters, such as betatron phase, while keeping the Twiss parameters matched (so called "phase-trombone" in accelerator jargon). The underlying logic is outlined in Section 6 above, while detailed formulation remains to be developed. Figure B12 illustrates the application of this principle to a realistic phase-trombone scenario.

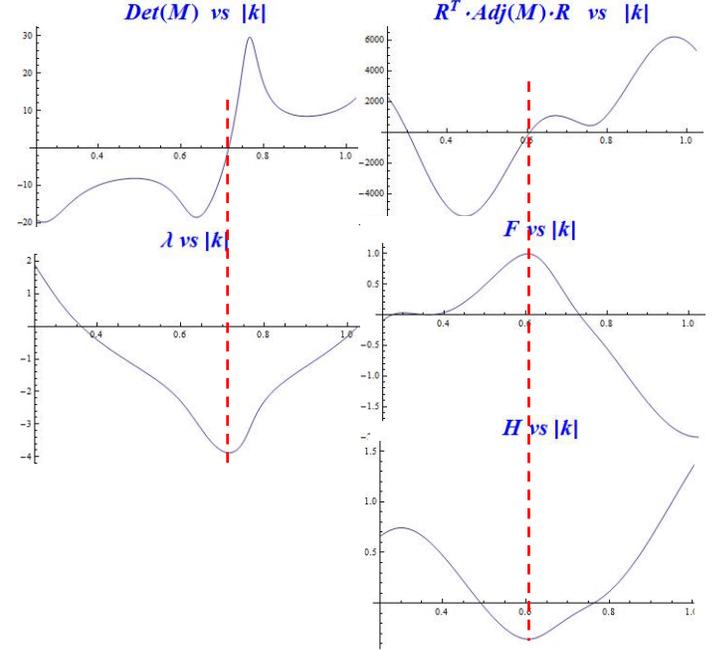

Figure B6. A numerical example showing evolution of objectives $F$ and $H$, and $\lambda$, extrema of which caused singularities in (B6) and (B8) as shown by 0 crossings in top plots. When $|k|$ is used as integration parameter instead, all quantities are well-behaved.

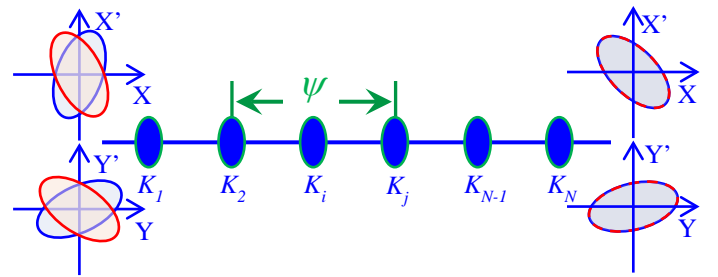

Figure B7. Basic configuration used for comprehensive numerical testing of the matching algorithm.

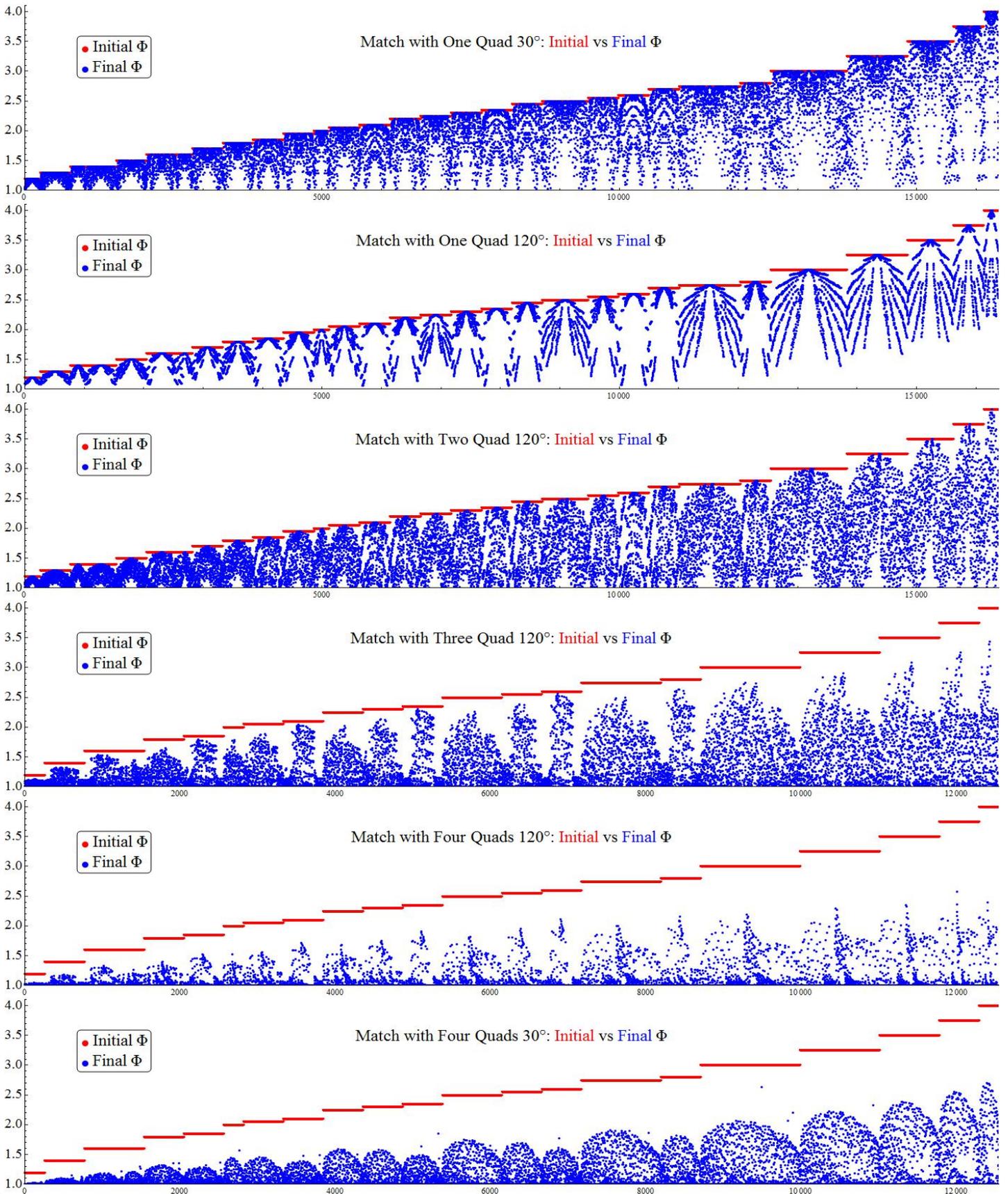

Figure B8. Matching using 1, 2, 3 and 4 quadrupoles in a 120° lattice. Also included are 1 & 4 quadrupoles at 30°. A total of over 15000 cases. Initial mismatch $\Phi$ covers a range 1.2–4.0, with $\Theta_{X/Y}$ covering 0–180°. They are ordered by initial $\Phi$ (red), for each of which there are many $\Theta_{X/Y}$. Blue dots correspond to optimal solution (smallest $\Phi$) reachable within monotonic trade-off. These data already form a prototype of the underline{interpolation table} of Section IV3, containing information needed to address underline{any} mismatch up to $\Phi=4$ via distributed matching.

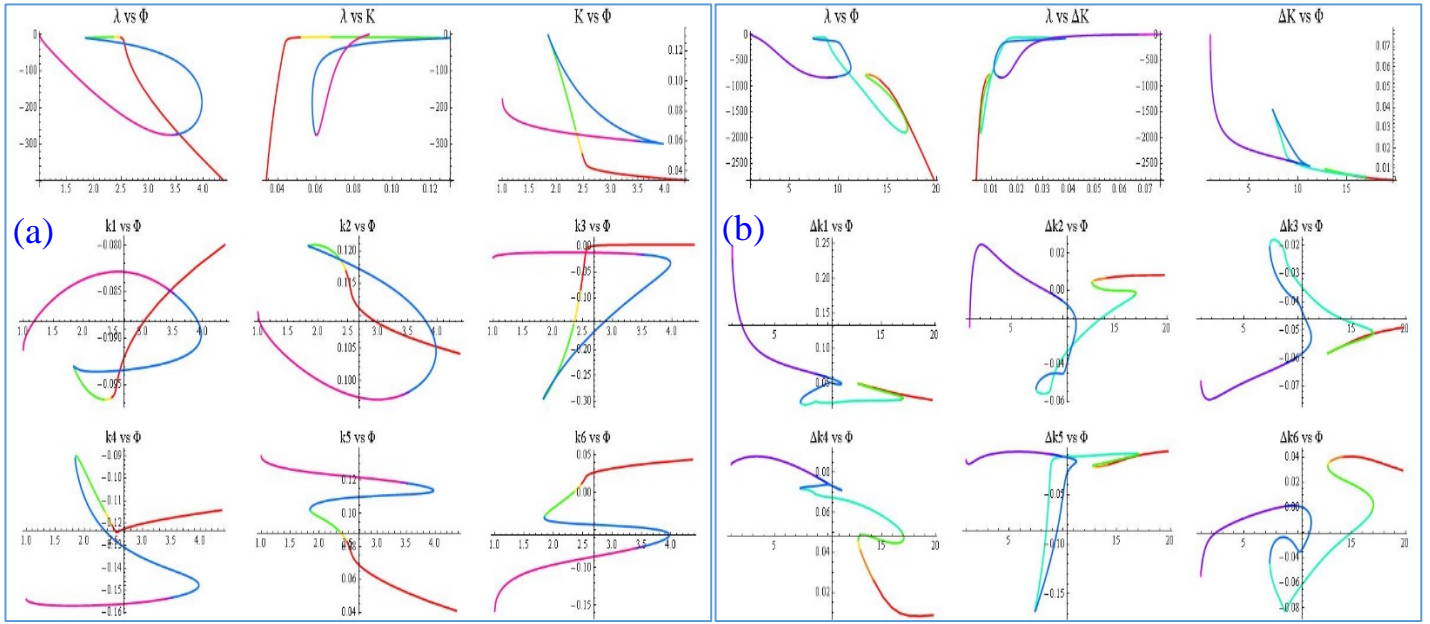

Figure B9. Paths followed by trade-off integration (from red to purple) in a 6-quadrupole 30° FODO lattice with embedded error. (a): Minimizing K. (b): Minimizing $\Delta K$. Each graph contains: Top row: Last part of the trade-off curve in projected spaces between $\lambda$–$\Phi$–K. Next two rows: Evolution of quadrupole $k_m$ or $\Delta k_m$, $m=1-6$, color correlated to top row. Despite convoluted patterns, Pareto fronts extracted from these curves are much simpler. For example $k_3$ in (a) remains close to 0 almost all the way (except small section) after Pareto front isolation.

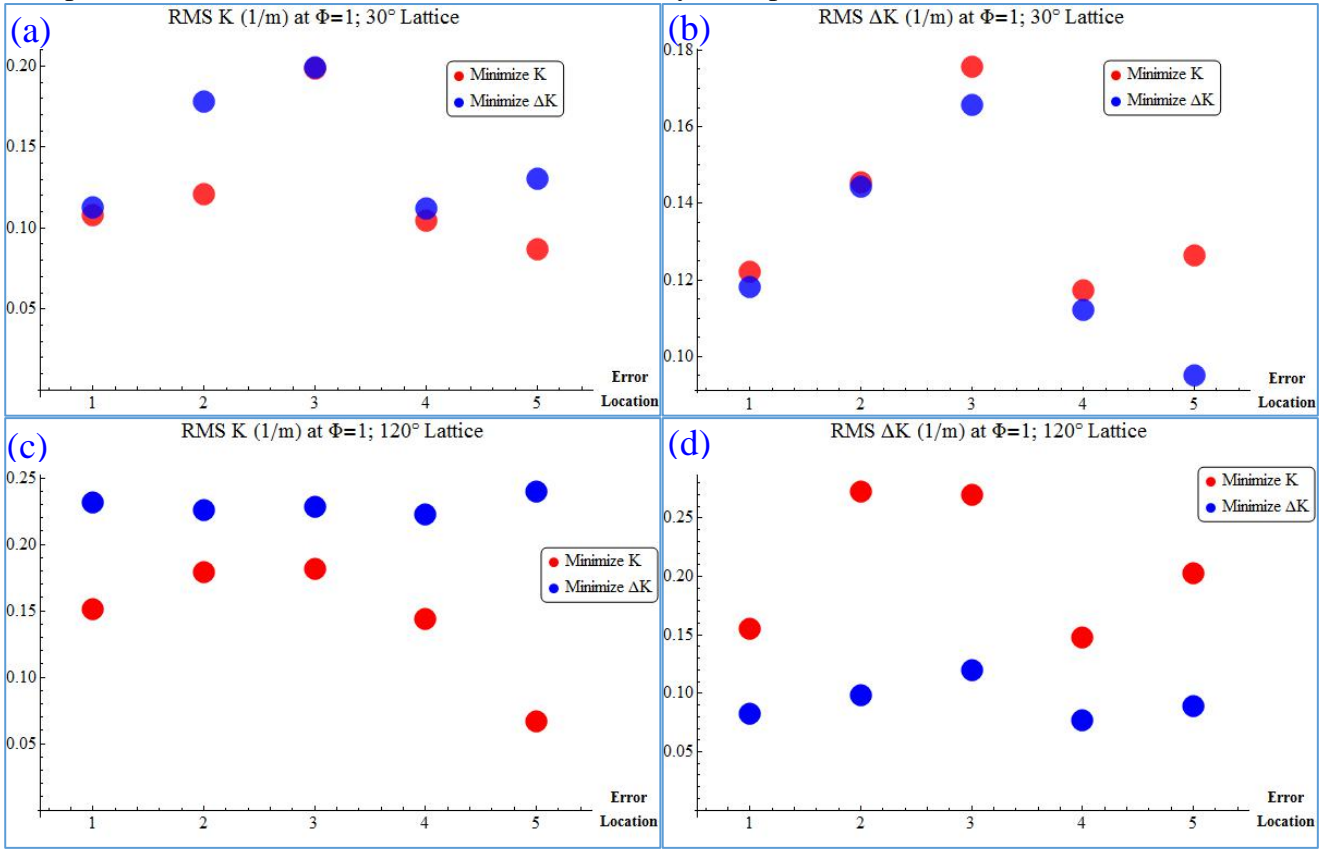

Figure B10. Final quadrupole RMS absolute strength K, or deviation $\Delta K$ from design, required to correct a fixed transport error embedded inside a 6 quadrupole section at one of the 5 possible intervals (horizontal axis). All dots correspond to final fully matched solution, $\Phi=1$, under different competing objectives. Red (Blue) dots correspond to solutions selecting K ($\Delta K$) as the competing objective in the trade-off integration. The final solution shows bias in favor of whichever competing objective is used, more pronounced in the 120° case. This is a desired result of the algorithm, biasing the solution toward user-selected constraints.

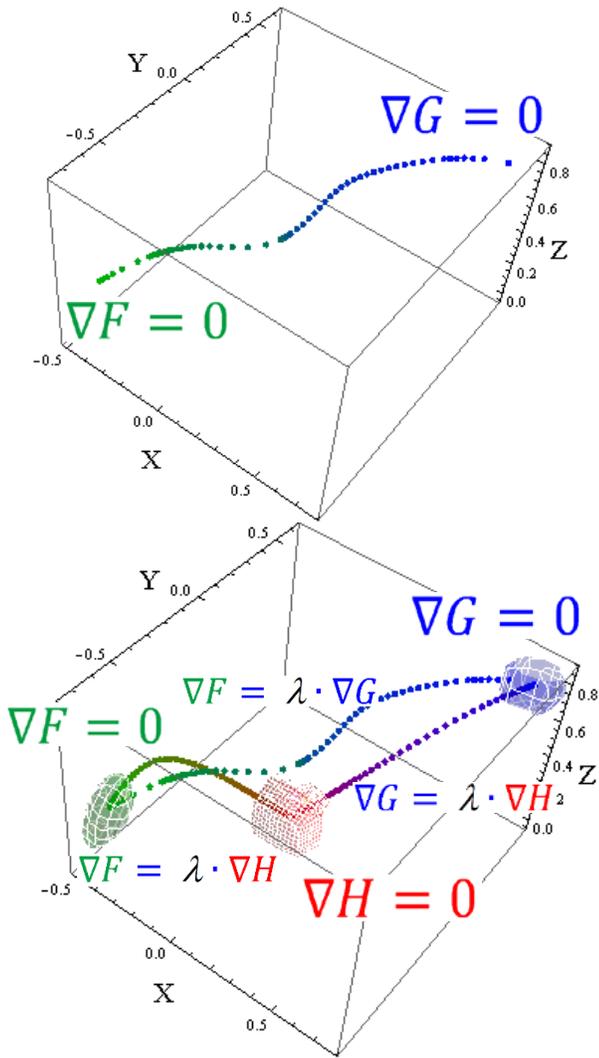

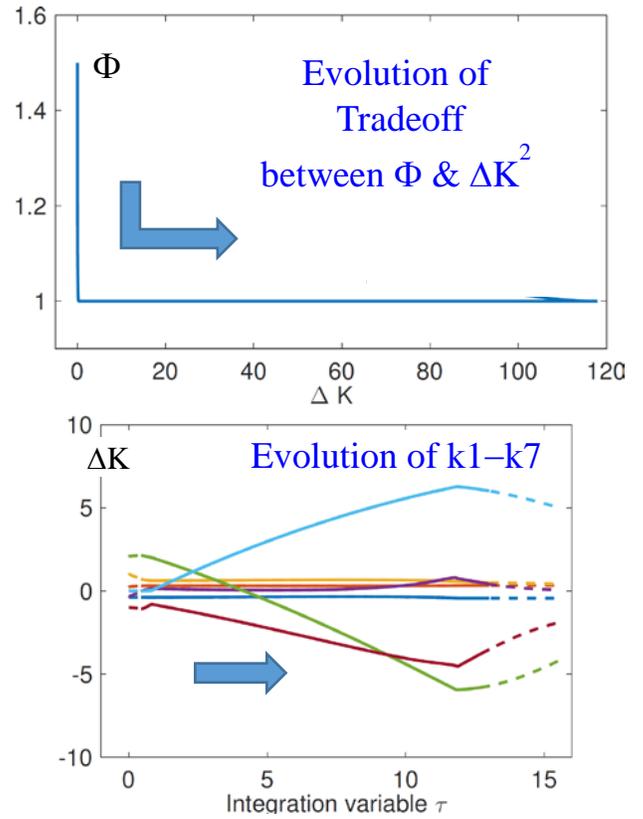

Figure B11. Restoring determinism in trade-off integration with no a priori knowledge of initial point. Top: Unknown trade-off curve between functions $F$ and $G$ of $(x, y, z)$, with no a priori knowledge of any point on it. Bottom: Artificial function $H$ with known initial point $\nabla H = 0$ is introduced, from which trade-off integration leads to $\nabla F = 0$, and onwards to $\nabla G = 0$. Equipotential surfaces for $F$, $G$ and $H$ at finite values are also shown.

Figure B12. Trade-off curve between $\Phi$ and $\Delta K^2$ (top) and evolution of $\Delta K$ for all quadrupoles (bottom) from applying the deterministic algorithm to an under-constrained matching system in a future design of LCLS-II optical lattice (Courtesy W. Qin). In this system the algorithm was used on 7 quadrupoles to satisfy 6 betatron matching constraints, leaving one degree of freedom to be explored. After reaching $\Phi=1$ in the tradeoff graph, the algorithm stays on $\Phi=1$ while advancing into regions of varying $\Delta K$ for all 7 quadrupoles, sampling continuously changing phase advance in the matched sections.